\documentclass[final,3p,times,twocolumn,authoryear,square]{elsarticle}
\pdfoutput=1
\usepackage{natbib}
\usepackage{amssymb}
\usepackage{amsthm}
\usepackage[squaren]{SIunits}
\usepackage{xspace}
\usepackage{bm}
\usepackage{multirow}
\usepackage{graphicx}% Include figure files
\usepackage{dcolumn}% Align table columns on decimal point
\usepackage{bm}% bold math
\usepackage[mathlines]{lineno}% Enable numbering of text and display math
\usepackage{amsmath}
\usepackage{amsbsy}
\usepackage{amsfonts}
\usepackage{amscd}
\usepackage{amstext}
\usepackage{subfig}

\newcommand{\tnu}{\tilde{\nu}_t} % turbulent viscosity in Spalart-Allmaras model

\usepackage[table]{xcolor}

\journal{European Journal of Mechanics-B/Fluids}

\include{macros}

\begin{document}

\begin{frontmatter}


%\title{Assessment of a curvature-sensitized Spalart-Allmaras turbulence model in incompressible flows over wavy walls}
%\title{A numerical approach for computing the evolution of incipient wind-generated waves over a viscous liquid}
\title{A combined VOF-RANS approach for studying the evolution of incipient wind-generated waves over a viscous liquid}

\author[inst1]{Florent Burdairon}
\ead{florent.burdairon@imft.fr}
 \affiliation[inst1]{organization={Institut de M\'ecanique des Fluides de Toulouse (IMFT)},%Department and Organization
            addressline={Universit\'e de Toulouse, CNRS}, 
            city={Toulouse},
            country={France}}
 \author[inst1]{Jacques Magnaudet\corref{cor}}
\cortext[cor]{Corresponding author}
\ead{jmagnaud@imft.fr} 
           
          %\footnote{{Email address for correspondence: julien.sebilleau@imft.fr}}

\begin{abstract}
Recent laboratory experiments have revealed that important insights into the physical processes involved in the wind-driven generation of surface waves may be obtained by varying the viscosity of the carrying liquid over several orders of magnitude. The present paper reports on the development of a companion approach aimed at studying similar phenomena through numerical simulation, a way expected to remove some of the experimental limitations, especially in the near-interface region, and to allow the relative influence of several physical processes to be assessed by disregarding or inactivating arbitrarily some of them. After reviewing available options, we select and approach based on the combination of a volume of fluid technique to track the evolution of the air-liquid interface, and a two-dimensional Reynolds-averaged version of the Navier-Stokes equations supplemented with a turbulence model to predict the velocity and pressure fields in both fluids. We examine the formal and physical frameworks in which such a time-dependent two-dimensional formulation is meaningful, and close the governing momentum equations with the one-equation Spalart-Allmaras model which directly solves a transport equation for the eddy viscosity. For this purpose, we assume the interface to behave as a rigid wall with respect to turbulent fluctuations in the air, and implement a versatile algorithm to compute the local distance to the interface whatever its shape. We first assess the performance of this model in the single-phase configurations of unseparated and separated flows over a wavy rigid wall, which are of specific relevance with respect to wind-wave generation. Then, we discuss the initialization protocol used in two-phase simulations, which involves an impulse disturbance with a white noise distribution applied to the interface position. We finally present some examples of interface evolutions obtained at several wind speeds with liquids of various viscosities, and discuss the underlying physics revealed by the associated statistics of interface disturbances, streamline patterns and energy spectra. 

\end{abstract}


%\begin{figure}

 %   \hspace{10mm}
  %  \includegraphics[width=0.8\textwidth]{Figures/L_100_Re_1600_tplus_5063.pdf}
  % \end{figure}



\end{frontmatter}


\section{Introduction}
\label{intro}
Understanding and predicting how wind blowing over a liquid generates waves at its surface and how the wave field in turn alters the turbulent motions above that surface and possibly below it has challenged oceanographers and fluid dynamicists for more than a century \citep{Komen1994, Jones2001, Janssen2004}. The manner in which turbulence influences or even governs the physical mechanisms involved in the wind-wave generation process is still in debate, although the two founding theories that attempt to rationalize these mechanisms are now two-thirds of a century old. In the first of them, Phillips \citep{Phillips1957} explored the possibility of a resonance between the turbulent pressure fluctuations in the boundary layer above the surface and the free deformation modes of that surface. This mechanism leads to a linear growth of surface deformations, and subsequent measurements have suggested that it may be appropriate for describing the very early stages of the wind-wave growth \citep{Kahma1988}. Simultaneously, Miles \citep{Miles1957} developed an inviscid theory in which surface deformations grow according to a two-dimensional linear instability mechanism leading to an exponential growth. In the original version of this theory, the sole role of turbulence is to set a logarithmic mean velocity profile in the boundary layer, which results in a mean shear stress at the liquid surface. Somewhat later, Miles improved his initial model in several respects, considering among other factors the influence of viscous corrections in the air flow \citep{Miles1959}, and that of the surface viscous stress in slightly viscous liquids  \citep{Miles1962}, which he showed to be significant in the generation of capillary and short gravity waves. Later, initially stimulated by Lighthill's reinterpretation of Miles' generic instability mechanism in terms of a `vortex force' \citep{Lighthill1962}, several attempts were carried out to include the influence of wave-induced turbulent stresses on the energy transfer from the air flow to the waves  \citep{Miles1967,Jacobs1987,vanDuin1992,Miles1993,Belcher1993,Miles1996}. Turbulent stresses modelled thanks to various closures were shown to increase significantly this transfer. The predicted growth rates display a reasonable, albeit closure-dependent, agreement with experimental data for short waves, but invariably under-predict dramatically the growth of long waves.  \\
 \indent For obvious reasons, most of the theoretical developments and experimental measurements to date have focused on the air-water system. However, considering more viscous liquids is also of the utmost interest. First of all, it provides a stringent test to available theoretical models. For instance, the minimum wind speed beyond which waves start to emerge and propagate increases strongly with the liquid viscosity \citep{Francis1954,Gottifredi1970,Paquier2016}, and a consistent model has to predict properly this increase. Also, it was recently shown that specific waves taking the form of `viscous solitons' develop at the surface of  liquids with a viscosity several hundreds times that of water \citep{Aulnette2019}. Understanding the underlying mechanisms and how the transition from regular wave trains to such viscous solitons operates is of clear interest from the point of view of pattern formation in free-surface flows \citep{Aulnette2022}. From a methodological viewpoint, one may expect viscous liquids to be better candidates than water to test the predictions of two-dimensional theories because entrainment of fluid particles by the air flow is more difficult in such liquids, making small-scale three-dimensional turbulence-driven motions less prone to develop at their surface. \\
\indent Up to now, the few studies devoted to the generation of wind waves at the surface of viscous liquids have been experimental in nature. High-precision optical techniques have been instrumental in the detection of minute interface deformations \citep{Moisy2009}, and particle image velocimetry has provided detailed access to the instantaneous velocity fields in vertical planes, both in the air and in the liquid, with the exception of the two-phase region located between the troughs and crests of the deforming interface. In air-water systems, numerical simulation has proven to efficiently complement laboratory measurements performed to unravel details of the interaction between the two fluids, especially within the crucial two-phase region where accurate measurements can hardly be achieved. Here we report on the development and application of a similar approach in the case of wind-wave generation at the surface of liquids of arbitrary viscosity. The present paper primarily aims at presenting step by step the elaboration of the corresponding computational strategy. This makes it basically methodological in nature, providing a `proof of concept' rather than a physical discussion of results. An extensive presentation of these results and a detailed discussion of the underlying physical processes is deferred to a forthcoming publication.\\
\indent The paper is organized as follows. Section \ref{review} discusses the various computational strategies developed over the last decades, and for some of them over the last few years, to simulate the evolution of wind-induced surface waves. In \S\,\ref{frame}, we specify the modelling framework adopted in the present investigation, and discuss its potentialities and intrinsic limitations. The set of governing equations considered in this approach, including the turbulence model, is detailed in \S\,\ref{turbmol}. The numerical framework and the specific issues related to the use of this turbulence model in the two-phase flow configurations of interest here are discussed in \S\,\ref{techni}. Performances of the turbulence model in single-phase configurations directly relevant to wind-wave generation, especially the flow over a rigid wavy wall, are presented and analyzed in \S\,\ref{ptest}. Section \ref{2phase} finally considers the canonical two-phase configurations relevant to the problem of wind-wave generation over a viscous liquid. We detail the initialisation protocol, a crucial aspect in this problem, and present some typical flow evolutions. We summarize the main findings of this study in \S\,\ref{conclu}.

\section{Review of available modelling strategies}
\label{review}
%In contrast with the air-water system, for which DNS is nowadays a natural tool to investigate the wave growth, the case of much more viscous liquids is more challenging.
Simulating the evolution of wind-generated surface waves opposes major numerical difficulties. Indeed, this situation combines the need to follow the deformation (and possibly the topological changes) of a gas-liquid interface, with that of simulating a high-Reynolds-number turbulent air flow.
Most numerical attempts to date dealt with the first issue by using coordinate transformations thanks to which the time-evolving interface is mapped onto a plane. In general, the underlying transformation is only performed on the vertical coordinate, so that the three coordinates in the transformed space are no longer mutually orthogonal. For this reason, numerous additional terms arise in the momentum equations \citep{Fulgosi2003,Komori2010,Yang2010,Zonta2015,Hao2019,Li2022}. If one is only interested in the very early stages of the wave development, the boundary conditions to be satisfied at the interface, i.e. the no-penetration (or kinematic) condition, together with the continuity of velocities and stresses, the latter including the hydrostatic and capillary contributions, may be linearized and projected onto the undeformed interface, thus avoiding the above coordinate transformation \citep{Tsai1998,Lin2008}. Of course, both approaches assume that the interface remains single-valued, which makes situations involving breaking waves out of reach. Recently, another route started to be explored by considering the potentialities offered by the volume of fluid (VOF) approach and some of its variants \citep{Yang2018,Wu2021,Wu2022}. The VOF approach is routinely used in the context of two-phase flows involving drops and bubbles. In this formulation, a single set of governing equations is solved throughout the flow domain, the local fluid properties (i.e. density and viscosity) and the capillary force being determined by computing the time advancement of the local volume fraction of one of the fluids. The key advantage of this approach is that one no longer needs to make the grid evolve in order to adjust to the instantaneous position of the interface, which allows for the use of the Cartesian form of the governing equations. The difficulty is transferred to the robustness and accuracy of the numerical schemes that are required to \text{(i)} ensure volume conservation of each fluid and \text{(ii)} deal with very large gradients in the physical properties (hence, in the velocity gradients) in the interfacial region. Nevertheless, major progress has been achieved on these technical aspects over the last two decades. This is why this approach is adopted in the present work.  \\
\indent The second issue is that air flows capable of generating waves at a liquid surface are necessarily turbulent. In the air-water system, tiny three-dimensional surface deformations start to be observed when the wind velocity in the bulk exceeds approximately 1.5 m.s$^{-1}$ \citep{Kahma1988}, and the critical wind velocity beyond which regular two-dimensional waves emerge directly is close to 3.6 m.s$^{-1}$ \citep{Paquier2016}. More commonly, air flow characteristics relevant to wind-wave generation are expressed in terms of the friction velocity, $u^*$, defined as the square root of the shear stress (divided by the air density) at the interface. The above two thresholds correspond to $u^*\approx0.07$ m.s$^{-1}$ and $u^*\approx0.18$ m.s$^{-1}$, respectively. State-of-the-art direct numerical simulations \citep{Li2022,Wu2022} handle friction velocities of the order of $0.1$ m.s$^{-1}$, and this approach looks ideal to get detailed insight into the processes involved in the wave generation mechanism. Nevertheless, dealing with liquids significantly more viscous than water imposes much more stringent requirements. For instance, two-dimensional waves at the surface of a liquid a hundred times more viscous than water only form when the air velocity exceeds $\approx8$ m.s$^{-1}$, i.e. the critical friction velocity is  approximately $u^*=0.35$ m.s$^{-1}$. Hence, there is a factor of 4 in between the critical conditions for such a system compared with the air-water system, and this translates into a factor of $4^{9/4}\approx23$ in the number of grid points required to solve the entire range of turbulent motions in the air, down to the Kolmogorov scale. Although such large simulations are not out of reach nowadays, they remain extremely expensive and do not allow a parametric study of the problem at a reasonable cost. \\
\indent An appealing alternative is to turn to large eddy simulation (LES). This approach has been extensively used by several groups in connection with the coordinate transformation technique discussed above. Such phase-resolved LES have been carried out over prescribed waves, be they periodic \cite{Sullivan2008,Zhang2019}, or distributed in the form of a broadband spectrum \cite{Sullivan2014} or a wave packet \cite{Sullivan2018}. In the most recent studies, LES is employed to solve the turbulent air flow, and a potential flow solver is used to evolve the wave train over time, with a surface pressure distribution exported from the LES field at the corresponding instant of time \cite{Hao2019}. A recent review of the various numerical techniques and subgrid-scale models employed in the LES approach, and of the advances it has provided in the understanding of wind-wave couplings is given in \cite{Deskos2021}. This review points out that LES has not yet been employed in conjunction with the VOF approach. The reasons are not discussed but are easy to understand. Indeed, modeling subgrid-scale transfers in regions alternately filled with liquid and air is a formidable task since, compared to single-phase turbulent flows, the filtered momentum equations involve many different second- and third-order unknown correlations for which appropriate closure laws have to be formulated \citep{Labourasse2007}. \\
\indent A simpler route has been used for several decades in connection with air-water interfaces distorted by a prescribed two-dimensional periodic wave. It consists in using a Reynolds-averaged version of the Navier-Stokes equations expressed in a reference frame travelling with the wave, coupled with a phenomenological turbulence closure relating the relevant components of the Reynolds stress tensor to the local characteristics of the flow field. A hierarchy of turbulence models has been used for this purpose, from the simplest zero-equation mixing length model \citep{Mastenbroek1996}, to the most sophisticated five-equation Reynolds-stress models \citep{Mastenbroek1996, Meirink2000, Li2000}, \textit{via} one-equation models combining a transport equation for the turbulent kinetic energy with a prescribed distribution for the turbulent integral length scale \cite{Gent1976, Li2000}. An important issue encountered with this approach stands in the boundary conditions at the interface. While each Reynolds-averaged velocity component is assumed to match the corresponding orbital velocity component at the water surface, phenomenological boundary conditions are employed for the turbulent unknowns. In the above references, these conditions rely on the existence of a logarithmic velocity profile and a local equilibrium between turbulent energy production and dissipation in the logarithmic region above the interface. This `wall-function' approach has the definite advantage of reducing the overall computational cost by avoiding the need for highly refined grids very close to the interface, and it appears suitable for dealing with already well-developed surface waves. Conversely, it constitutes one of the main limitations of the Reynolds-averaged approach as soon as the prediction of the evolution of small-amplitude interface deformations is concerned, since the local characteristics and the longitudinal profiles of the turbulent stresses in the viscous sublayer and the buffer layer above the interface are then expected to play a major role \cite{Miles1993,Belcher1993,Belcher1998}. Therefore, in such a context, turbulence models allowing the use of `natural' boundary conditions for the turbulent quantities right at the interface are mandatory.
%Here we explore a different route which consists in considering that the air-liquid interface deforms over time but the characteristics of the turbulent air flow that are relevant with respect to the fate of these deformations may be predicted using a time-dependent Reynolds-averaged approach.
\section{Modelling framework}
\label{frame}
In what follows, we make use of the above Reynolds-averaged framework to represent the effects of turbulence above the air-liquid interface and possibly below it, in conjunction with a turbulence model obeying a `natural' condition at the interface, the evolution of which is tracked with a VOF approach. Examining the evolution of non-periodic time-dependent interface deformations while representing effects of turbulence through a Reynolds-averaged approach may seem contradictory at first glance. The crucial underlying issue is that of the separation between turbulent and orbital velocity and pressure fluctuations, which is an extremely complex task in the presence of a non-periodic three-dimensional wave field  \citep{Benilov1970,Thais1995,Hristov1998}. Here, `orbital' refers to fluctuations resulting from or correlated with any normal displacement of the interface. Obviously, the above issue simplifies drastically if a clear separation of time scales between the two components exists, as the response of waves evolving over `long' time scales to turbulent fluctuations covering a range of `short' time scales may then be studied in the framework of governing equations averaged over an intermediate time scale. However, this appealing framework is unfortunately not appropriate, since the characteristic time scales of wind-generated interface deformations generally overlap those involved in the turbulent motion \citep{vanDuin1992}. Therefore, a strict and consistent separation can only be achieved by introducing a drastic simplifying assumption.\\
\indent In what follows, we assume that interface displacements are two-dimensional, taking place in the $(x,z)$ plane, say, while turbulent fluctuations are of course three-dimensional, i.e. they depend on both the local position ${\bf{X}}=(x,y,z)$ and time, $t$. With this assumption, any quantity $\Phi({\bf{X}},t)$ may be decomposed in the form 
\begin{equation}
\Phi({\bf{X}},t)=\langle{\Phi}\rangle ({\bf{x}},t)+\Phi'({\bf{X}},t)\,,
\end{equation}
where the operator $\langle\cdot\rangle$ corresponds to a spatial averaging in the spanwise direction ($y$), ${\bf{x}}=(x,z)$ is the projection of the local position onto the vertical plane, and $\Phi'({\bf{X}},t)$ stands for the turbulent fluctuation of $\Phi$. With this definition, $\langle{\Phi}\rangle$ and $\Phi'$ are uncorrelated, so that $\langle{\langle{\Phi}\rangle\Phi'}\rangle\equiv0$. The above assumption, already used by Miles \citep{Miles1967},  also implies that quantities directly related to the instantaneous position of the interface do not have a turbulent component. In the framework of the VOF approach, this position is defined with the help of the volume fraction of one of the fluids, say $C$. Geometrical properties of the interface, such as its local unit normal, $\bf{n}$, and mean curvature, $\kappa=\nabla\cdot\bf{n}$, are also defined using the first and second derivatives of $C$ in the form $\bf{n}=\nabla C/||\nabla C||$ and $\kappa=\nabla\cdot\bf{n}$, respectively. Similarly, the local density and viscosity of the two-fluid medium only depend on $C$ and possibly on its gradients. For the above reason, $C$ and the above geometrical or physical properties do not have a turbulent component. Hence, for instance, 
\begin{equation}
C({\bf{X}},t)\equiv\langle C\rangle({\bf{x}},t)\,,
\label{Cdef}
\end{equation}
 so that $\langle C\Phi\rangle=\langle \langle C\rangle\Phi\rangle= \langle C\rangle\langle\Phi\rangle$. The key advantage of the above assumption is that the averaged Navier-Stokes equations based on the application of the operator $\langle\cdot\rangle$ are similar to the classical incompressible Reynolds-averaged equations. In particular, turbulence only appears through the second-order correlation tensor $\langle\bf{u}'\bf{u}'\rangle$, with ${\bf{u}}({\bf{X}},t)=\langle{\bf{u}}\rangle({\bf{x}},t)+{\bf{u}}'({\bf{X}},t)$ the local fluid velocity. If needed, $x$-averaged quantities, hereinafter denoted with an overbar, may be defined by integrating  the quantity of interest over the appropriate distance, which has to be much larger than the lowest wavenumber present in the spectrum of $C$. These averaged quantities still depend on $z$ and $t$, e.g., $\overline{\Phi}(z,t)$, and the difference $\langle\Phi\rangle({\bf{x}},t)-\overline{\Phi}(z,t)$ represents the orbital contribution to $\Phi$.\\
 \indent Obviously, the above simplifying assumption reduces severely the generality of the physical situations that may be studied in the corresponding framework. In particular, it prevents any progress in the study of the development of the longitudinal streaks and three-dimensional tiny `wrinkles' that deform the air-water interface at low wind speeds and/or short fetches, and may be thought of as the precursors of two-dimensional waves \cite{Kahma1988,Caulliez1998,Veron2001,Paquier2016}. More globally, it removes any possibility of examining the relevance of the intrinsically three-dimensional Phillips mechanism %according to which waves grow due to a resonance between turbulent pressure fluctuations in the air and free deformation modes of the interface 
 \citep{Phillips1957}. Despite these severe restrictions, the proposed approach is appealing in that it is potentially suitable for exploring the two-dimensional evolution of the interface in connection with Miles instability mechanism \citep{Miles1957,Benjamin1959,Miles1959,Lighthill1962}. Indeed, this instability scenario considers the evolution of a two-dimensional interface $z=\eta(x,t)$ subjected to pressure and shear stress distributions resulting from a prescribed $\overline{{\bf{u}}}(z)$-velocity profile in the air flow, set by turbulent motions in the boundary layer. Turbulence models operating in the framework of the Reynolds-averaged momentum equations have precisely be designed to predict the corresponding mean shear. After Miles established his initial `quasi-laminar' theory \citep{Miles1957}, the role of wave-induced (or `orbital' according to the present terminology) turbulent stresses in the wave growth was reconsidered \citep{Townsend1972,Belcher1993} and found to overcome the efficiency of the original mechanism in the growth of young waves \citep{Ayet2022}. This influence may also been assessed in the framework defined above. This is why one can expect that this route may provide interesting new insights into the efficiency and relevance of the Miles mechanism and its variants in the generation of wind waves at the surface of liquids of arbitrary viscosity, provided a suitable turbulence model is employed.\\

%We consider incompressible turbulent flows in which a suitable averaging operator may be defined to separate mean quantities (denoted with an overbar) from fluctuating quantities (denoted with a prime). 
\section{Governing equations and turbulence model}
\label{turbmol}
In principle, a Reynolds stress model solving transport equations for the four components $\langle u'^2\rangle,\,\langle v'^2\rangle,\,\langle w'^2\rangle,\,\langle u'w'\rangle$ and for at least an extra scalar quantity related to the integral length scale (such as the dissipation rate) is desirable. Indeed, such models are designed to account for nonlocal and non-equilibrium effects which are expected to take place in the outer part of the boundary layer above the air-liquid surface when the latter deforms \citep{Belcher1993,Miles1996,Belcher1998}. However, designing proper near-interface modifications in the transport equation for the dissipation rate or any related scale-defining quantity is far from obvious, a difficulty most of the time circumvented through the use of a `wall-function' approach. For this reason, we did not retain this type of model here and rather opted for a much simpler one-equation eddy-viscosity model. As will be shown later, despite several deficiencies, this model accurately reproduces the mean flow profile and captures most of the important aspects of the air flow variations above a wavy surface in the regimes of interest here. Therefore, we consider it as a useful step to explore the potentialities of the general approach designed above, although more sophisticated models will certainly have to be considered in the future. \\
\indent Having selected an eddy-viscosity closure, the Reynolds stress tensor $\langle{{\bf{u'}\bf{ u'}}}\rangle$ is related to the strain-rate tensor $\langle{\bf{S}}\rangle=\frac{1}{2}\left(\nabla\langle{\bf{u}}\rangle+\nabla\langle{\bf{u}}\rangle^{\text{T}}\right)$ in the form $\langle{{\bf{u'}\bf{ u'}}}\rangle=\frac{2}{3}\langle k\rangle{\bf{I}}-2\nu_t\langle{\bf{S}}\rangle$, with $\langle k\rangle=\frac{1}{2}\langle{{\bf{u'}\cdot\bf{ u'}}}\rangle$ the turbulent kinetic energy per unit mass, $\nu_t$ the eddy viscosity and ${\bf{I}}$ the unit tensor. In the framework of the VOF approach, the Reynolds-averaged two-phase flow is then governed by the averaged Navier-Stokes equations
\begin{eqnarray}
\label{divu}
\partial_tC+\langle{\bf{u}}\rangle\cdot\nabla C&=&0\,,\\
    \nabla\cdot\langle{\bf{u}}\rangle&=&0\,,\\
    \nonumber
   \partial_t (\rho\langle{\bf{u}}\rangle)  +  \nabla\cdot (\rho\langle{\bf{u}}\rangle \, \langle{\bf{u}}\rangle )  &=& \rho{\bf{g}}\\
  - \nabla\mathcal P   + 2\nabla\cdot [ (\mu+\rho\nu_t)\, \langle{\bf{S}}\rangle] &-&\gamma\kappa\nabla C\,,
    \label{eq:Re}
\end{eqnarray}
with $\rho$ and $\mu$ the local density and viscosity of the two-fluid medium ($\nu=\mu/\rho$ being the kinematic viscosity), $\gamma$ the surface tension, $\mathcal{P}=\langle{P}\rangle+\frac{2}{3}\rho \langle k\rangle$ the modified pressure, ${\bf{g}}$ denoting gravity. In \eqref{eq:Re}, the capillary force is expressed using the classical formalism introduced in \cite{Brackbill1992} and the mean curvature of the interface is computed as $\kappa=\nabla\cdot(\nabla C/||\nabla C||)$. The local density and viscosity are defined through the linear relations
\begin{equation}
\rho(C)=C\rho_l+(1-C)\rho_a,\quad\quad \mu(C)=C\mu_l+(1-C)\mu_a\,,
\end{equation}
with indices $l$ and $a$ referring to the properties of the liquid and air, respectively. \\%While the linear dependence of $\rho$ with respect to $C$ is exact, 
\indent Similar to all quantities in \eqref{divu}-\eqref{eq:Re}, the eddy viscosity is defined throughout the flow domain, so that no boundary condition can be imposed on it at the interface. Therefore, $\nu_t$-variations across the successive subregions of the boundary layer above the interface and possibly below it have to be directly obtained $via$ the turbulence model. This means that this model has to take into account in one way or another the distance to the interface to achieve the proper behaviour of the Reynolds stresses in its vicinity. In what follows, we consider that the air flow `feels' the interface as a rigid wall. More specifically, what is assumed here is that the normal velocity fluctuation is zero at the interface, so that the non-diagonal component of the Reynolds stress tensor ($\langle u'w'\rangle$ in the case of a flat interface located at a constant $z$) vanishes there. This assumption is relevant given that (i) turbulence originates in the air flow, not in the liquid, and (ii) the density ratio $\rho_a/\rho_l$ is very small, so that a normal velocity fluctuation in the air flow barely deforms the liquid surface. Obviously, the flow in the liquid may also be turbulent if the entrainment by the air flow is strong enough and the liquid has a low enough viscosity. This point will be discussed later.\\
\indent To predict the variations of the eddy viscosity, we selected the Spalart-Allmaras model \citep{SA1992_art}. This model, widely used in the context of high-Reynolds-number aerodynamic flows over complex geometries, directly solves a transport equation for the eddy viscosity, following the early proposal of \cite{Nee1969}. More precisely, in a single-phase wall-bounded flow, the model first determines an auxiliary turbulent viscosity, $\tnu$, by solving the transport equation
\begin{eqnarray}
   \label{SA}
     \partial_t \tnu&+&\langle{\bf{u}}\rangle\cdot\nabla\tnu =    c_{b1}f_r \tilde{\Omega} \tnu - c_{w1} f_w \left[ \frac{\tnu}{\ell} \right]^2 \\
     \nonumber
     &+& \frac{1}{\sigma} \bigg\{ \nabla \cdot [(\nu + \tnu) \nabla \tnu] + c_{b2} \nabla \tnu\cdot\nabla \tnu \bigg\}\,.
\end{eqnarray}
In (\ref{SA}), $\ell$ is the local distance to the wall, and $\tilde{\Omega}$ denotes a positive scalar quantity which, beyond the viscous and buffer regions, equals the local vorticity magnitude $\Omega=(2\,\boldsymbol{\Omega}\colon\boldsymbol{\Omega})^{1/2}$, with $\boldsymbol{\Omega}=\frac{1}{2}\left(\nabla\langle{\bf{u}}\rangle-\nabla\langle{\bf{u}}\rangle^{\text{T}}\right)$ the rotation-rate tensor. The various functions and constants in \eqref{SA} are determined in such a way that, in a near-wall region, $\tilde\nu_t\propto\ell$ and $\tilde{\Omega}\propto\ell^{-1}$ down to the wall; see \ref{SAM} for details. Then, the eddy viscosity $\nu_t$ involved in (\ref{eq:Re}) is related linearly to $\tnu$ \textit{via} an empirical damping function, $f_{v1}$, in the form \begin{equation}
\nu_t =  f_{v1}\tnu\,.
\label{nutilde}
\end{equation}
  The  function $f_{v1}$  tends to $0$ as the wall is approached and is unity far from it. This damping function allows accurate estimates of the turbulent shear stress to be obtained in near-wall regions, including the buffer and viscous sublayers. The various empirical functions and constants involved in \eqref{SA}-\eqref{nutilde} are detailed in \ref{SAM}. Interestingly, in \cite{Miles1993}, Miles suggested that the Spalart-Allmaras model, which was brand new at that time, could be a good candidate to explore the influence of the orbital Reynolds stresses on the growth of wind-generated waves.\\
  \indent The above model extends straightforwardly to the air flow involved in the two-phase configurations of interest here, provided $\ell({\bf{x}},t)$ is considered as the distance from any position located in the air flow, i.e. in the region such that $1-C({\bf{x}},t)>0.5$, to the interface. The evaluation of $\ell$ will be detailed in the next section. 
  % More specifically, what is assumed implicitly by leaving this model unmodified and considering that $\ell(x,z,t)$ is the local distance to the interface (the evaluation of which is detailed below) is that the normal velocity fluctuation is zero at the interface, so that the non-diagonal component of the Reynolds stress tensor ($\langle u'w'\rangle$ in the case of a flat interface located at a constant $z$) vanishes there. This assumption is relevant in the present context, given that the density ratio $\rho_a/\rho_l$ is very small. Similar to all quantities in \eqref{divu}-\eqref{eq:Re}, the eddy viscosity is defined throughout the flow domain, so that no boundary condition can be imposed to \eqref{SA} at the interface. 
  Variations of $\nu_t$ across the boundary layer are obtained through the empirical functions $f_w$ and $f_{v1}$ involved in \eqref{SA}-\eqref{nutilde}, plus the near-wall correction of the vorticity magnitude $\tilde{\Omega}$ (see \eqref{omegatilde}). \\
  \indent If the flow in the liquid beneath the interface is considered laminar, as it will be in %{\color{red}{most of}} 
  the examples discussed in \S\,\ref{evol2}, this condition is straightforwardly enforced by extending \eqref{nutilde} throughout the two-phase flow domain in the form
  \begin{equation}
  \nu_t =  (1-C)f_{v1}\tilde\nu_{t>}\,,
  \label{SAm}
  \end{equation}
  where $\tilde\nu_{t>}({\bf{x}},t)$ denotes the solution of \eqref{SA} computed when the considered position ${\bf{x}}$ stands in the air at time $t$. The `constitutive' equation \eqref{SAm} leaves the eddy viscosity in the air flow unchanged but sets it to zero in the liquid. In cases where the flow in the liquid is turbulent, as happens with water beyond wind speeds of a few meters per second, the model may easily be adapted to predict the eddy viscosity in the liquid. Indeed, for physical reasons discussed in \ref{turbliq}, turbulence beneath an air-liquid interface behaves differently from that close to a rigid wall. %For this, it must be kept in mind that the subsurface flow is submitted to a prescribed shear at the interface. For this reason, provided the interface is uncontaminated by surfactants, 
  %In particular, tangential velocity fluctuations do not generally vanish at the interface. %on it, while the normal fluctuation does if turbulence is `not too strong', i.e. if it leaves the surface undeformed. Existence of non-vanishing tangential fluctuations implies that 
 As a result, the turbulent shear stress in the liquid grows linearly with the distance to the interface, even within the viscous sublayer, just as $\tilde\nu_t$ does in \eqref{SA}. Consequently, one merely needs 
 %the Spalart-Allmaras may be straightforwardly adapted to predict 
 to set $f_{v1}\equiv1$ in \eqref{nutilde} to obtain a realistic decay of the eddy viscosity as the interface is approached from below. In such turbulent-turbulent configurations, the eddy viscosity throughout the two-phase flow becomes
  \begin{equation}
  \nu_t =  (1-C)f_{v1}\tilde\nu_{t>}+C\tilde\nu_{t<}\,,
  \label{SAD}
  \end{equation}
with $\tilde\nu_{t<}({\bf{x}},t)$ the local value of $\tnu$ computed from \eqref{SA} when the considered position stands in the liquid, i.e. $C({\bf{x}},t)>0.5$.

\section{Numerical framework and specific techniques}
\label{techni}
The governing equations (\ref{divu})-(\ref{SA}) are solved using the JADIM code developed at IMFT. This second-order finite volume code solves the Navier-Stokes equations and scalar transport equations on a staggered grid. Equations are written in general orthogonal curvilinear coordinates \citep{Magnaudet1995}, which makes the treatment of the curved geometries considered in some test cases of \S\,\ref{ptest} straightforward. %This functionality will be used below to analyze the performance of the turbulence model in the case of a single-phase flow above a rigid wavy wall. 
Time-advancement is achieved with a third-order Runge-Kutta scheme for advective and source terms, and a semi-implicit Crank-Nicolson scheme for viscous terms \citep{Calmet1997}. Incompressibility is enforced to machine accuracy at the end of each time step by solving a Poisson equation for the pressure increment. Equation \eqref{divu} governing the evolution of the interface is solved using a flux-limiting transport scheme split into a succession of one-dimensional steps \citep{Bonometti2007}. In (\ref{SA}), the production term and the nonlinear diffusion term, $\sigma^{-1} c_{b2}\nabla \tnu\cdot\nabla \tnu$, are treated as source terms, while the Fick-like diffusion term, $\sigma^{-1} \nabla \cdot [(\nu + \tnu) \nabla \tnu]$, and the wall-destruction term are handled with the Crank-Nicolson scheme; treating this destruction term implicitly contributes to the numerical stability of the overall algorithm.\\
\indent In all computations reported below, the flow takes place over a rigid wall on which Dirichlet conditions 
\begin{equation}
{\bf{\langle{u}\rangle}}={\bf{0}}\,,\quad \tnu=0\,,
\label{dirich}
\end{equation}
 are imposed. The upper boundary may either be a symmetry plane or a rigid flat wall. In the former case, free-slip conditions 
 \begin{equation}
 {\bf{\langle{u}\rangle}}\cdot{\bf{n}_u}=0\,,\, {\bf{n}_u}\cdot\nabla[{\bf{\langle{u}\rangle}}-({\bf{\langle{u}\rangle}}\cdot{\bf{n}_u})\,{\bf{n}_u}]={\bf{0}}\,,\, {\bf{n}_u}\cdot\nabla\tnu=0
 \label{free}
 \end{equation}
  are imposed, ${\bf{n}}_u$ denoting the unit normal to the considered plane. The flow is periodic in the streamwise $(x)$ direction in all cases. %However, tests not described here were also carried out in the backward-facing step configuration, using inflow and non-reflecting outflow conditions. 
{%\color{green}{Initial conditions will be specified in due course for each configuration.}}
\\
\indent An important issue in the determination of the turbulent viscosity is the evaluation of the distance $\ell({\bf{x}},t)$ to the interface at a given position ${\bf{x}}$. In the case of a single-phase wall-bounded flow, we compute $\ell$ by first building an explicit parametric representation of the wall geometry in the form ${\bf{x}}_w(s)\equiv(x_w(s),z_w(s))$ with $ds=(dx_w^2+dz_w^2)^{1/2}$ the infinitesimal arc-length element. Then we vary $s$ along the wall (actually within a `reasonable' finite interval of streamwise positions $x_w$ on both sides of $x$), compute the distance $\ell_s({\bf{x}})=\{(x-x_w(s))^2+(z-z_w(s))^2\}^{1/2}$ within this interval and set $\ell({\bf{x}})=\text{min}|_{s}(\ell_s({\bf{x}}))$. In two-phase configurations, assuming that the interface remains single-valued, we locate it using the standard SLIC technique \citep{Noh1976}. That is, starting from the top of the domain, we examine the volume fraction in the successive grid cells encountered at a given $x$ by decreasing $z$ until we detect the first cell in which $C({\bf{x}},t)>0.5$. This gives a first approximation of the interface position at the considered abscissa, $z_i(x,t)$. Then, the precise position of the interface, $z_s(x,t)$, is obtained by requesting that the liquid volume enclosed below the interface equals that given by the volume fraction field, which yields 
\begin{equation}
z_s(x,t)=z_i(x,t)+\int_{z_i}^{z_s}C(x,z',t)dz'\,.
\label{slic}
\end{equation}
 We then compute $\ell({\bf{x}},t)$ as in the single-phase case. However, it may happen, especially in long runs, that the interface does not remain single-valued, since some wave components may overturn. In such cases, the above approach is no longer sufficient, as it only allows the uppermost position of the interface at a given abscissa to be detected. To deal with these more general situations, we employ the following strategy. We define a disc of radius $R$ centered at the current location ${\bf{x}}$, and introduce the index $\alpha(x',z')$ such that $\alpha=1$ at all cell positions $(x',z')$ belonging to the disc, i.e. such that $(x'-x)^2+(z'-z)^2\leq R^2$, and $\alpha=0$ otherwise. Then, starting with the initial guess $R=\ell({\bf{x}},t)$ provided by the above SLIC technique, we compute the average volume fraction within the disc, i.e. 
 \begin{equation}
 \mathfrak{C}(R,t)=\frac{\int_{\mathcal{V}}\alpha(x',z')C(x',z',t)d\mathcal{V}}{\int_{\mathcal{V}}\alpha(x',z')d\mathcal{V}}\,.
 \label{disc}
 \end{equation}
  We compare $ \mathfrak{C}(R,t)$ with the local volume fraction at the disc centre,  $C({\bf{x}},t)$. If the relative difference between the two exceeds a prescribed tolerance, this is an indication that the volume fraction has varied within the disc, so that the distance to the interface is actually less than $R$. Therefore, we decrease $R$ until $ \mathfrak{C}(R,t)$ becomes close enough to $C({\bf{x}},t)$. The final determination of $\ell({\bf{x}},t)$ is obtained by interpolating the last two values of $R$ involved in the iterative process, as the final one underestimates the distance to the interface while the penultimate one overestimates it.
\section{Preliminary tests: single-phase computations}
\label{ptest}
%\subsection{Fully-developed half-channel flow}
Before considering wind-wave generation, it is necessary to assess the performances and limitations of the turbulence model in relevant single-phase flows. The simplest configuration of  interest here is presumably the fully-developed turbulent flow in a plane channel. This flow is also relevant with respect to the initialization of two-phase configurations because the corresponding velocity and turbulent viscosity fields are used during this stage, as will be seen in \S\,\ref{2phase}. The corresponding tests are detailed in \ref{channel}. It is shown that the model allows the mean velocity profile to be predicted accurately, even at high Reynolds number, with a very limited number of grid points located in the viscous sublayer. This is a good indication that the various near-wall and viscous corrections involved in \eqref{SA}-\eqref{nutilde} perform well in a zero-pressure gradient flow over a flat wall.\\
\indent A more complex single-phase configuration directly relevant to wind-wave generation is that of a turbulent flow over a rigid wavy wall. This configuration has been extensively studied experimentally by Hanratty and coworkers, both in unseparated and separated configurations resulting from small- \citep{Zilker1977,Thorsness1978,Frederick1988} and large- \cite{Zilker1979,Buckles1984,Kuzan1989} amplitude undulations, respectively. The turbulence response to these undulations revealed several subtleties, starting with the amplitude and phase shift of the streamwise variations of the wall pressure and shear stress, and the position of the separation and reattachment points in the separated case. Measurements in the unseparated configuration established that standard mixing length-type models are unable to predict the above phase shifts for `long' undulations with wavenumber $k$ such that $k^+=k\delta_\nu\lesssim0.01$, with $\delta_\nu=\nu/u^*$ the near-wall viscous length scale. In contrast, accurate predictions were obtained whatever $k^+$ by introducing an \textit{ad hoc} `relaxation' of the mixing length accounting for the nonlocal influence of the streamwise pressure gradient on the thickness of the buffer layer and viscous sublayer \citep{Abrams1985}. Since then, this flow configuration has been extensively investigated experimentally  (with large-amplitude undulations) \cite{Kruse2003,Wagner2007,Hamed2015,Segunda2018} and computationally, be it through DNS \citep{Maass1996,DeAngelis1997,Cherukat1998,Yoon2009}, LES \citep{Henn1999,Cui2003,Chang2004,Wagner2011}, or RANS simulations with two-equation turbulence models \citep{Patel1991, Chang2004, Knotek2012, Chaib2015,Segunda2016, Segunda2018}.\\
\indent Here we assess the performance of the Spalart-Allmaras model in two distinct wavy wall configurations, corresponding to unseparated and separated flows, respectively. These two cases, based on the experimental conditions of \citep{Zilker1977,Frederick1988} on the one hand and \cite{Buckles1984} on the other hand, were taken as reference in a previous LES study \citep{Henn1999}, and these LES predictions are also used below for the sake of comparison. The flow domain is a wavy channel with total height $H$ from the wave trough ($z=-a$) to an upper rigid flat wall ($z=H-a$), both walls been subjected to a no-slip condition. The shape of the wavy wall is defined as $z_w(x)=a\cos(2\pi x/\lambda)$ and the wave steepness $2a/\lambda$ is 0.031 in the unseparated case and 0.2 in the separated one, the wavelength being close to the mean channel height $H_m=H-a$ in both cases. The flow Reynolds number $Re_b$ based on the height $H_m/2$ and bulk velocity $U_b=H_{ref}^{-1}\int_{0}^{H_m} \overline{u}(z')dz'$ is 6560 in the unseparated case (with $H_{ref}=H_m$) and 10600 in the separated one (with $H_{ref}=H_m-a$). The grid is curvilinear and orthogonal,  with 144 cells from wall to wall and 52 cells over one wavelength. It is nonuniform across the channel, with a minimum near-wall cell size $\Delta z_{min}^+=\Delta z_{min}\overline{u^*}/\nu$ close to $0.04$ and 0.09 in the unseparated and separated cases, respectively, $\overline{u^*}$ denoting the wavelength-averaged friction velocity on the wavy wall. Computations are carried out by prescribing the pressure gradient $d\overline{\mathcal{P}}/dx$ and initializing the velocity and turbulent viscosity fields from $z=0$ to $z=H_m$ with the corresponding Poiseuille profile and the parabolic distribution $\tnu({\bf{x}},t=0)=40\,\nu z(H_m-z)/H_m^2$, respectively. In addition, the initial velocity and turbulent viscosity distributions are both set to zero in the trough region $-a\leq z\leq 0$.\\
\begin{figure*}[h!]
    \center
    \hspace{0mm} \includegraphics[width=0.33\textwidth]{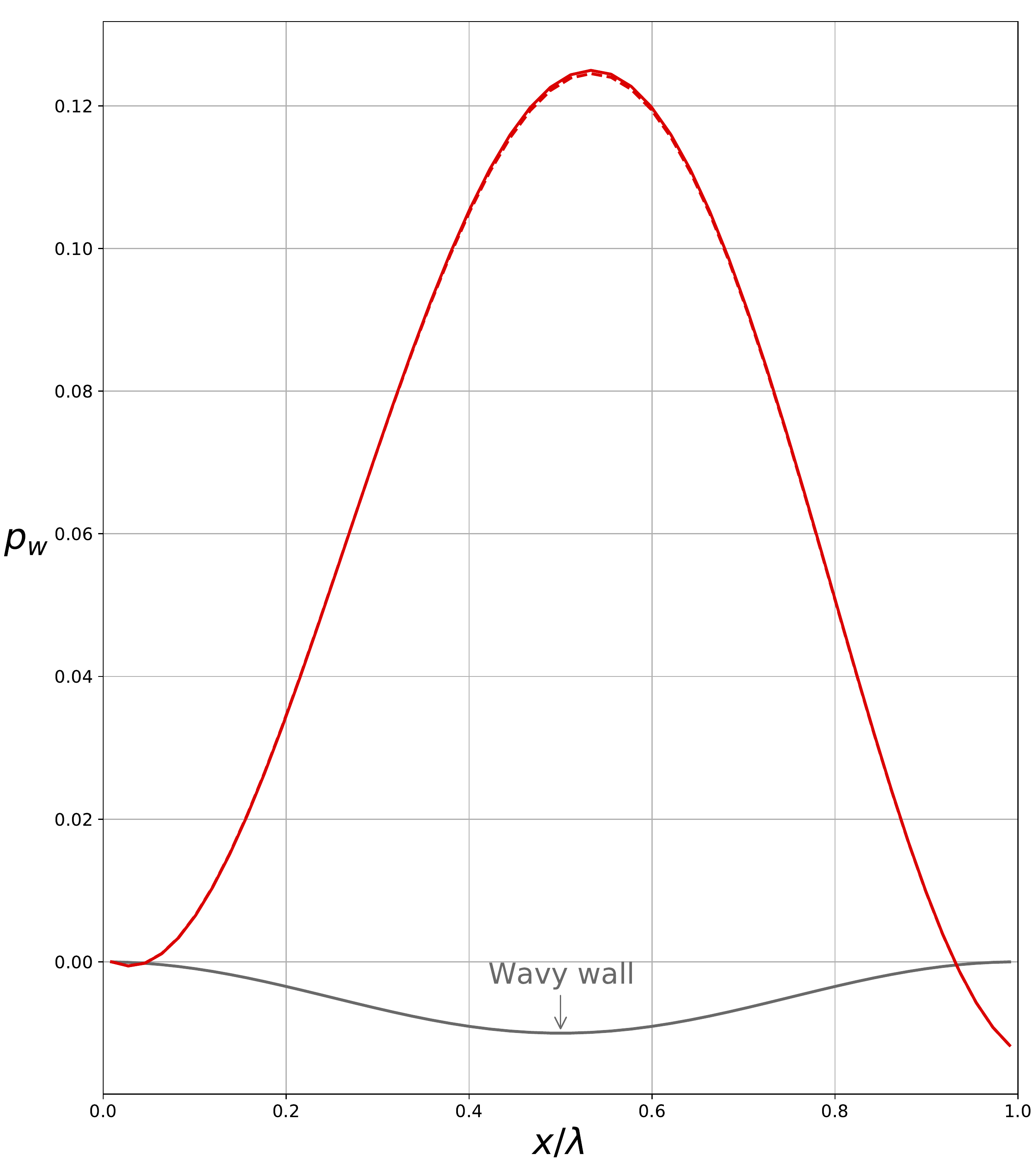}
   %    \vspace{-1mm}\hspace{10mm}$(a)$
     \hspace{5mm}\includegraphics[width=0.33\textwidth]{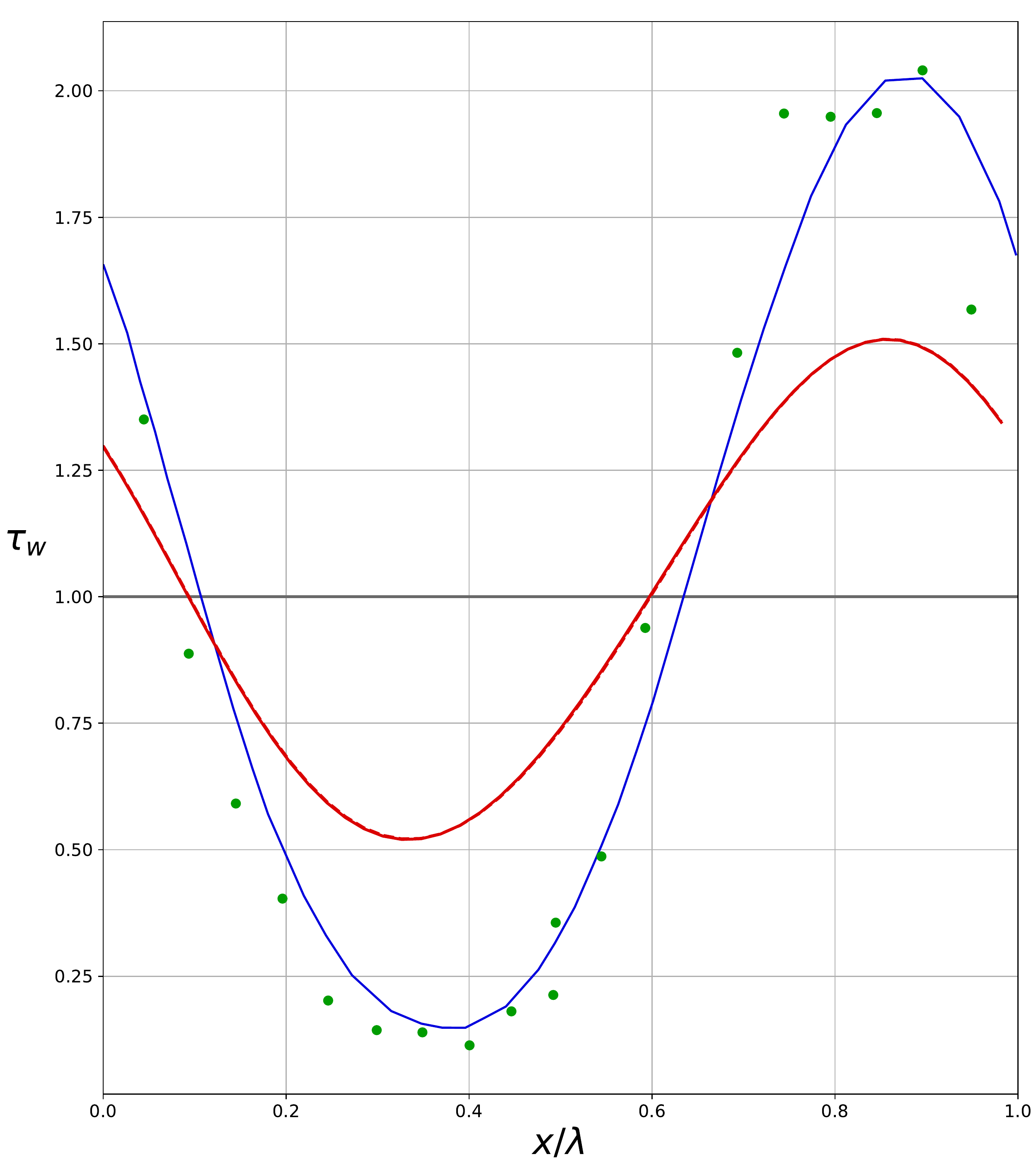}\\
 \vspace{-2mm}\hspace{-40mm}$(a)$\hspace{55mm}$(b)$
    \caption{Wall distributions in a non-separated flow over a wavy wall with $2a/\lambda\approx0.031$, $H_m/\lambda=1$ and $Re_b\approx6560$ ($Re^*=H_m\overline{u^*}/2\nu=370$, $k^+=\pi Re^{*-1}H_m/\lambda\approx8.5\times10^{-3}$). $(a)$: normalized pressure $p_w$; $(b)$: normalized shear stress. {\color{red}{-----}}: present predictions with the standard turbulence model ($f_r=1$ in (\ref{SA})); {\color{red}{- - - -}}: predictions with the curvature-sensitized turbulence model ($f_r$ given by \ref{fr})-(\ref{constr})); {\color{green}{$\bullet$}}: experiments \citep{Zilker1977} ($Re_b\approx7850$); {\color{blue}{-----}}: LES prediction \citep{Henn1999}.}
      \label{nonsep}
\end{figure*}
\begin{figure*}[h!]
    \center
    \hspace{0mm} \includegraphics[width=0.33\textwidth]{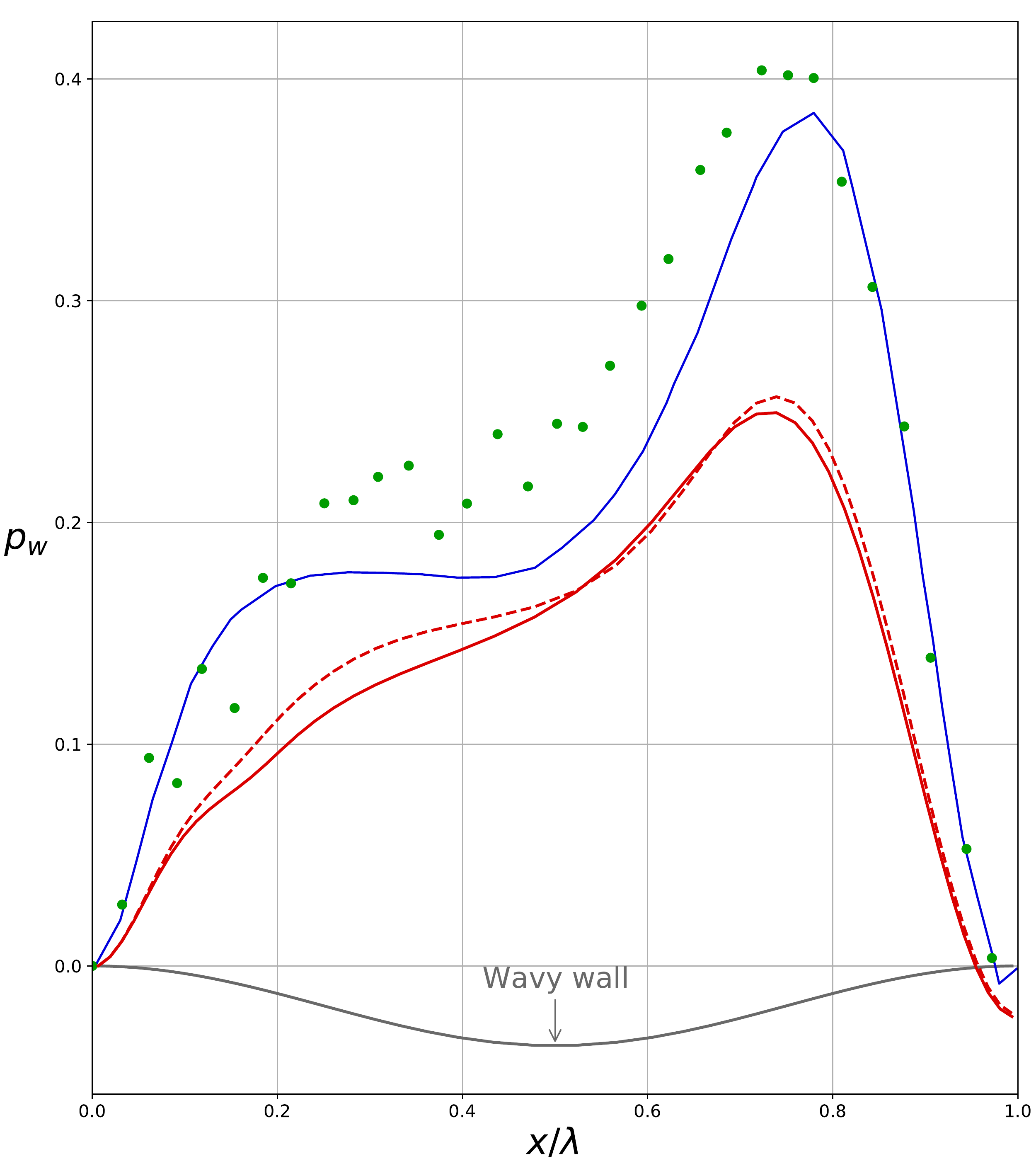}
   %    \vspace{-1mm}\hspace{10mm}$(a)$
     \hspace{5mm}\includegraphics[width=0.33\textwidth]{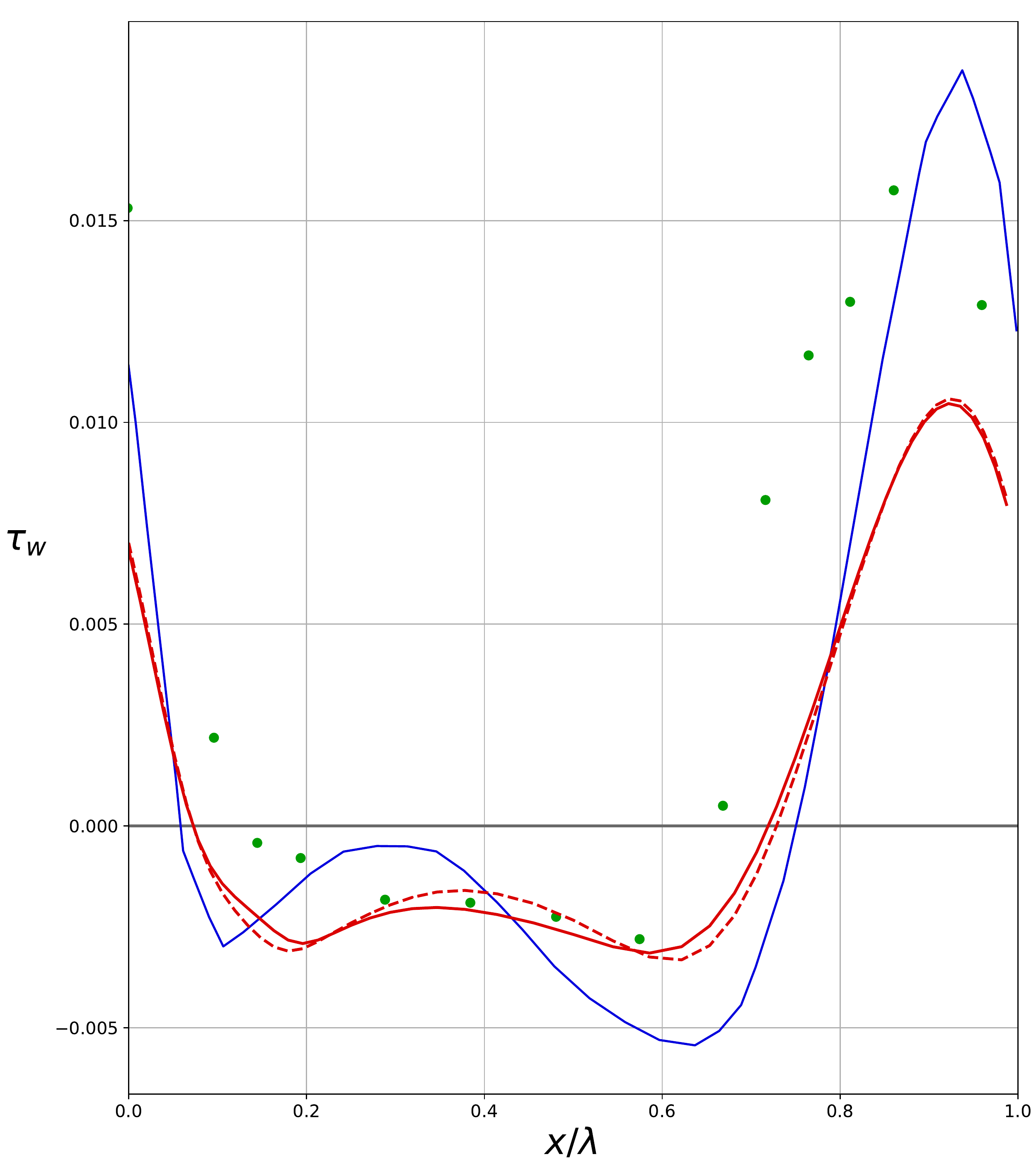}\\
 \vspace{-2mm}\hspace{-40mm}$(a)$\hspace{55mm}$(b)$
    \caption{Wall distributions in a separated flow over a wavy wall with $Re_b\approx10600$ ($Re^*=1370$), $H_m/\lambda=0.933$ and $2a/\lambda=0.2$.  For caption, see figure \ref{nonsep}; experimental data are taken from \citep{Buckles1984} ($Re_b\approx12000$).}
      \label{sep}
\end{figure*}
\begin{figure*}[h!]
    \center
    \hspace{0mm} \includegraphics[width=0.45\textwidth]{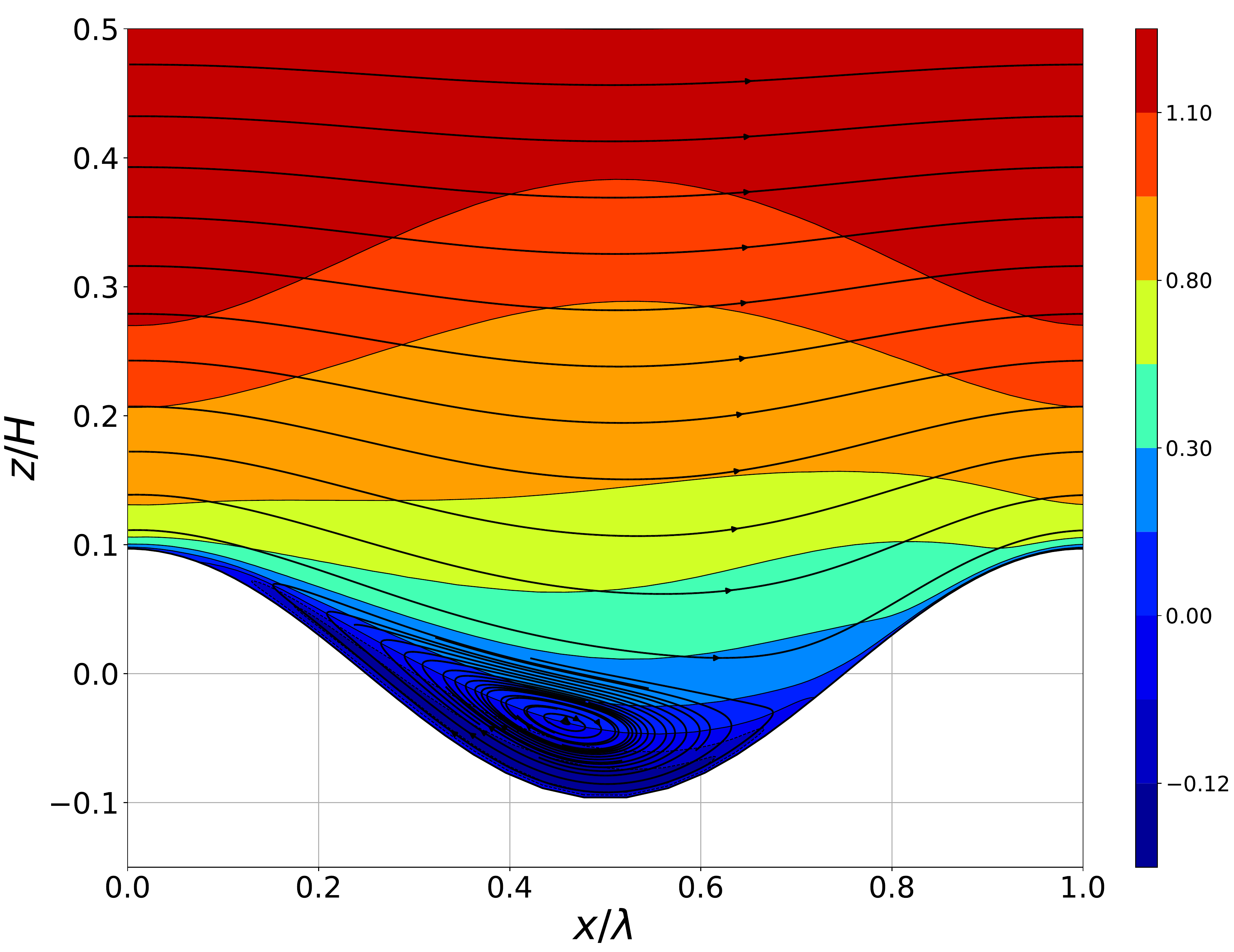}
   %    \vspace{-1mm}\hspace{10mm}$(a)$
     \hspace{5mm}
     \includegraphics[width=0.45\textwidth]{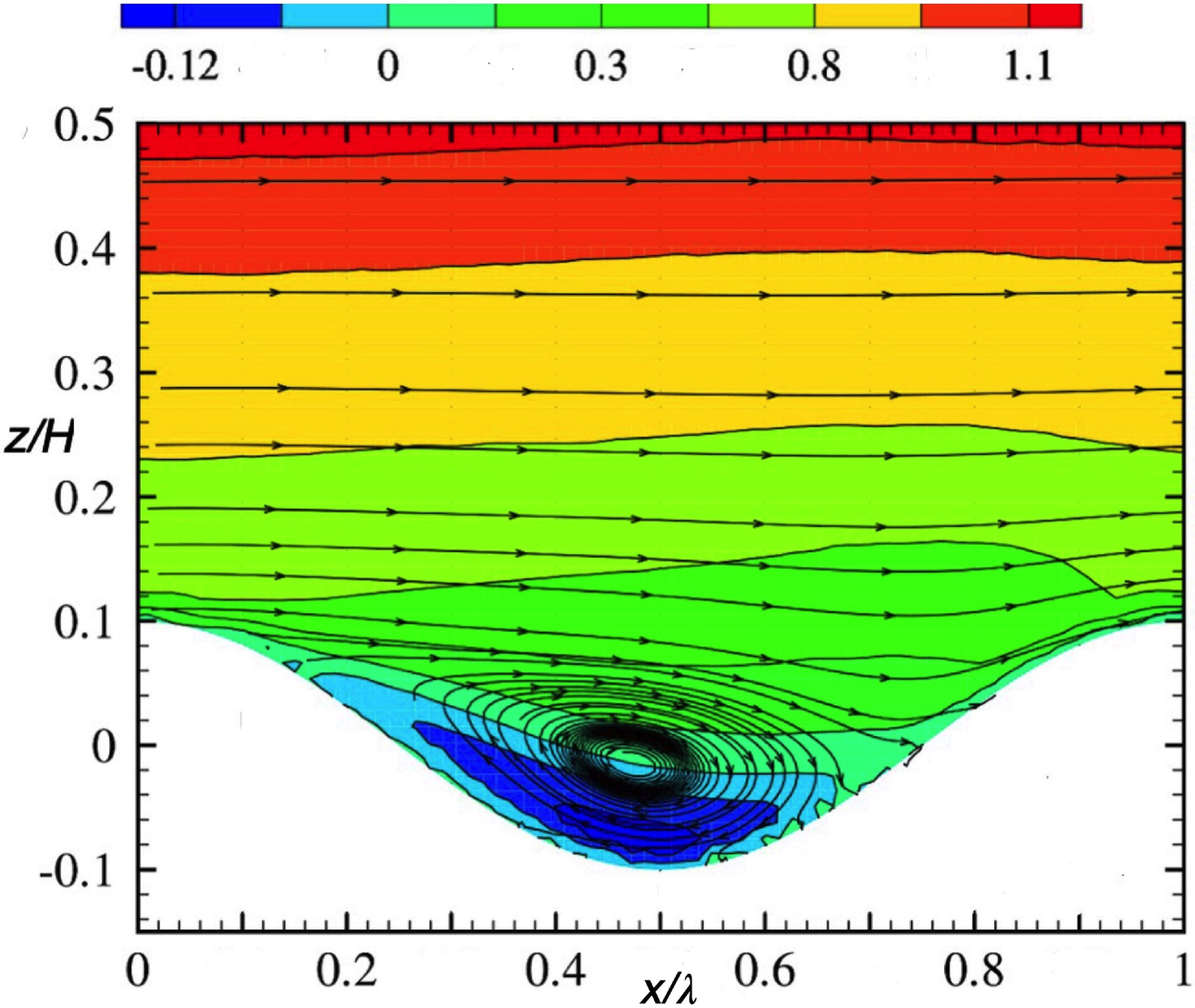}\\
 \vspace{-2mm}\hspace{-50mm}$(a)$\hspace{78mm}$(b)$
    \caption{Streamines in a separated flow over a wavy wall with $Re_b\approx10600$, $H_m/\lambda=0.933$ and $2a/\lambda=0.2$. $(a)$: present results; $(b)$: experiments (adapted from figure 5$(b)$ of \cite{Segunda2018}). The color scale corresponds to isovalues of the streamwise mean velocity $\langle u\rangle(x,z)$.}
      \label{segunda}
\end{figure*}
Figure \ref{nonsep} displays the pressure and shear stress distributions along the wavy wall in the unseparated case. The dimensionless wall pressure and shear stress are respectively defined as $p_w(x/\lambda)=(\mathcal{P}(x,z_w(x))-\mathcal{P}(0,z_w(0))/(\frac{1}{2}\rho U_b^2)$ and $\tau_w(x/\lambda)=\mu\partial_n\langle u\rangle(x,z_w(x))/(\rho\overline{u^*}^2)$, with $\partial_n$ the normal derivative with respect to the wall. In line with experiments at a higher $Re_b$ \citep{Zilker1977}, the wall pressure predicted by the one-equation turbulence model (subfigure $(a)$) is seen to exhibit a nearly harmonic response, with a maximum located slightly downstream of the trough. In contrast, the shear stress profile (subfigure $(b)$) is shifted ahead of the wall profile by nearly $55^\circ$. As the figure shows, this phase shift is in good agreement with experimental results and LES predictions. However, the magnitude of the shear-stress variations is severely under-estimated, especially on the wind-ward side of the crest where the model predicts a $50\%$ increase of the shear stress with respect to its mean value, while the experimental data and the LES result both indicate a twofold increase.\\
\indent The distributions of $p_w$ and $\tau_w$ in the separated case are shown in figure \ref{sep}. Here again, the general appearance of the two distributions is correctly captured but the maxima, located $90^\circ$ and $50^\circ$ ahead of the crest for $p_w$ and $\tau_w$, respectively, are under-estimated by nearly $30\%$. The wall shear stress is predicted to become negative at $x/\lambda\approx0.072$ and to return to positive at $x/\lambda\approx0.720$, in slightly better agreement with experimental measurements than the LES prediction. Nevertheless, experiments rather indicate that the flow detaches at $x/\lambda\approx0.13$ and reattaches at $x/\lambda\approx0.66$, so that the model actually somewhat anticipates the detachment on the leeward side of the crest and slightly delays the reattachment. Second-order models systematically suffer from the same shortcoming \cite{Segunda2016}. The model predicts that the position of the $p_w$-maximum coincides with that of the reattachment point, in line with the experimental findings of \cite{Zilker1979}. In the detached region, the wall shear stress predicted by the Spalart-Allmaras model is in better agreement with the nearly flat distribution revealed by experiments than that computed in the reference LES. Predictions of the curvature-sensitized version of the model, in which the magnitude of the production term in (\ref{SA}) is modulated by the local ratio of the rotation and strain rates according to \eqref{fr}, exhibits marginal differences with those of the standard model. The most noticeable difference is found on the wall pressure, which increases on the leeward side of the crest when curvature effects are accounted for, making a `shoulder' appear in the $p_w$-distribution, in line with that exhibited by experimental and LES results. \\
\indent Figure \ref{segunda} shows the streamline pattern in the separated case, and compares it with the experimental pattern determined in \cite{Segunda2018}. The separated zone is seen to extend over most of the region located above the trough in both panels. Nevertheless, the recirculation region predicted with the Spalart-Allmaras model is flatter than that observed in the experiment. For instance, in the plane $x/\lambda=0.5$, i.e. right above the through, the top of the recirculation stands at the altitude $z/H\approx0$, while it is detected at $z/H\approx0.06$ in the experiment. Also, the footprint of the disturbance induced in the distribution of the streamwise velocity $\langle u\rangle(x,z)$ by the wavy wall persists deeper in the bulk in the computation. Indeed, a significant bump, associated with a velocity minimum, is noticed in the core region $(z/H\lesssim0.5)$, while the corresponding experimental distribution exhibits a much flatter profile. \\
\indent In summary, the tests performed in an unseparated turbulent flow over a wavy wall reveal that the one-equation Spalart-Allmaras model correctly predicts the phase of the wall pressure and shear stress distributions in the considered case, which corresponds to $k^+\approx8.5\times10^{-3}$. Additional results for other $k^+$ are reported in \ref{Phase}. They confirm these conclusions for larger $k^+$ (relevant to wind-wave generation), while for smaller $k^+$ (more relevant to swell propagation) the model fails to predict correctly the phase lag, as systematically observed with turbulence models based on the eddy viscosity concept \citep{Abrams1985, Belcher1993,Belcher1998}. In the separated configuration, the model  predicts the position of the detachment and reattachment points fairly well, although it under-predicts the extent of the recirculating region in the direction normal to the wall. In both cases, the main deficiency of the model with respect to present purposes appears to be the significant under-prediction of the amplitude of the $p_w$- and $\tau_w$-variations, especially that of their peak values. In other terms, the Spalart-Allmaras model seems to `soften' flow variations along the sinusoidal wall profile too much. It might be that this deficiency can be attenuated by tuning some of the empirical functions and constants of the models detailed in \eqref{Cornu}-\eqref{const}. However, these parameters have been calibrated in a number of flows in the past, and any change in one of them improving the predictions in the specific configuration considered here might be detrimental in other flows. This is why we did not attempt to tune any of these parameters. The only attempt we made consisted in setting the diffusion term (last term in the right-hand side of \eqref{SA}) to zero in the streamwise direction, to favor sharper variations of the eddy viscosity along the flow. However, this attempt did not reveal any significant change in the $p_w$- and $\tau_w$-distributions, nor in those of the mean velocity field $\langle u\rangle(x,z)$.

\section{Two-phase configurations}
\label{2phase}
\subsection{Geometry and grid}
We now turn to the two-phase computations based on the complete set of equations \eqref{divu}-\eqref{SA}. The simulations are carried out in a rectangular domain with dimensions $L\times H$ in the streamwise $(x)$ and vertical $(z)$ directions, respectively. The flow is assumed periodic in the $x$-direction, while no-slip and free-slip conditions apply on the bottom and top walls, respectively. In most runs, the mean level of the interface, $z_0$, is assumed to stand midway between the bottom and upper surfaces, which is close to the experimental conditions of \cite{Paquier2016}. Computations make use of a grid with a nonuniform cell spacing in the $z$-direction on both sides of the mean interface. More precisely, the cell spacing decreases gradually from $z=0$ to $z=z_0-\Delta/2$, stays constant within the stripe $z_0-\Delta/2\leq z\leq z_0+\Delta/2$, and re-increases from $z=z_0+\Delta/2$ to the upper free-slip wall located at $z=H$. We set the grid spacing in the intermediate stripe to $\Delta z_{min}\approx0.5\delta_\nu$, with $\delta_\nu=\nu/u^*$ the characteristic near-wall length scale of the air flow. The thickness of the stripe is set to $\Delta/H=1/8$, which, under most conditions, makes the refined region thick enough to track the possible development of surface waves until saturation. This choice yields a number of cells $N_\Delta=\frac{1}{2}Re^*$ within this stripe. The lower and upper non-uniform regions are discretized with $0.35N_\Delta$ cells each, so that the height of the largest cells (adjacent to the bottom wall and upper free-slip surface, respectively), is approximately $10\delta_\nu$. A uniform cell spacing, $\Delta x$,  is imposed in the streamwise direction. To avoid excessive cell distortion in the refined region, the ratio $\Delta x/\Delta z_{min}$ is maintained close to $5$. This yields a total number of cells in the streamwise direction $N_x\approx0.8 Re^*L/H$. Gathering the above information, the total number of cells is seen to be approximately $0.8 Re^*L/H\times0.85Re^*\approx0.7(L/H)Re^{*2}$. For instance, with $Re^*=800$ and $L/H=3/2$, the grid comprises a total number of $1026\times678$ cells.
\subsection{Initialization protocol}
 \begin{figure*}[t!]
    \center
    \hspace{-7mm} \includegraphics[width=0.33\textwidth]{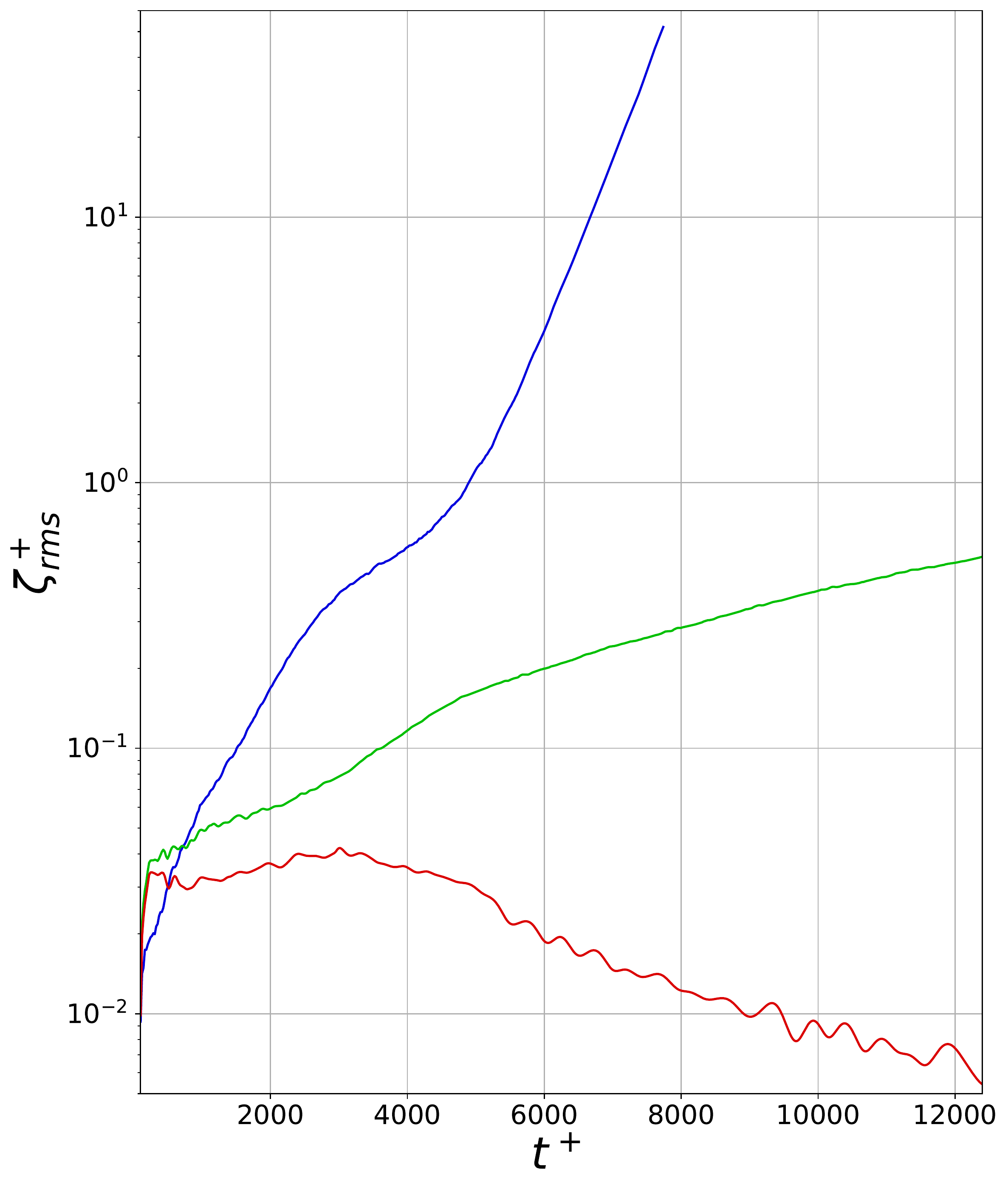}
   %    \vspace{-1mm}\hspace{10mm}$(a)$
     \hspace{-0.5mm}\includegraphics[width=0.33\textwidth]{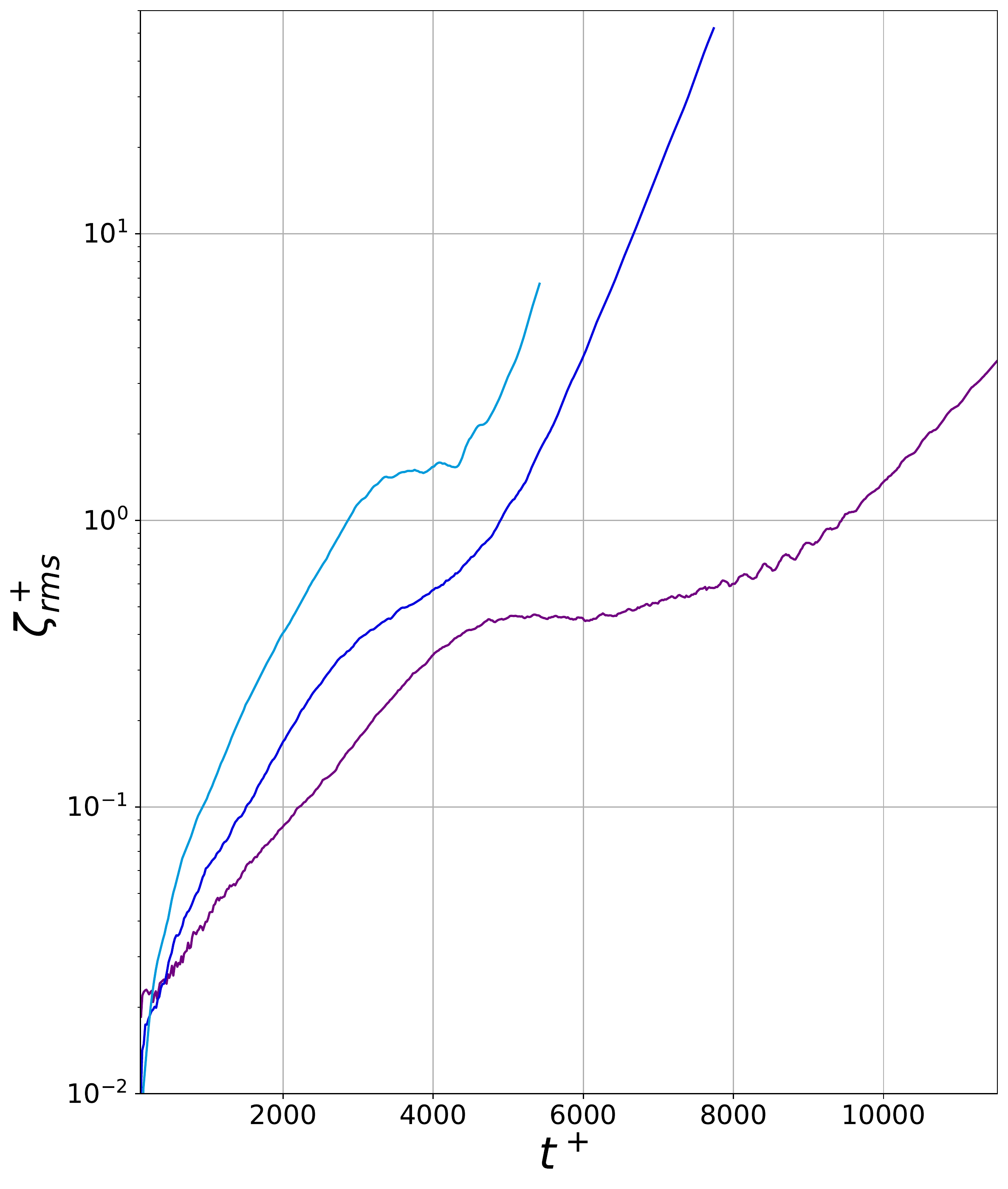}
  %    \vspace{-1mm}\hspace{10mm}$(b)$
     \hspace{-0.5mm}\ \includegraphics[width=0.33\textwidth]{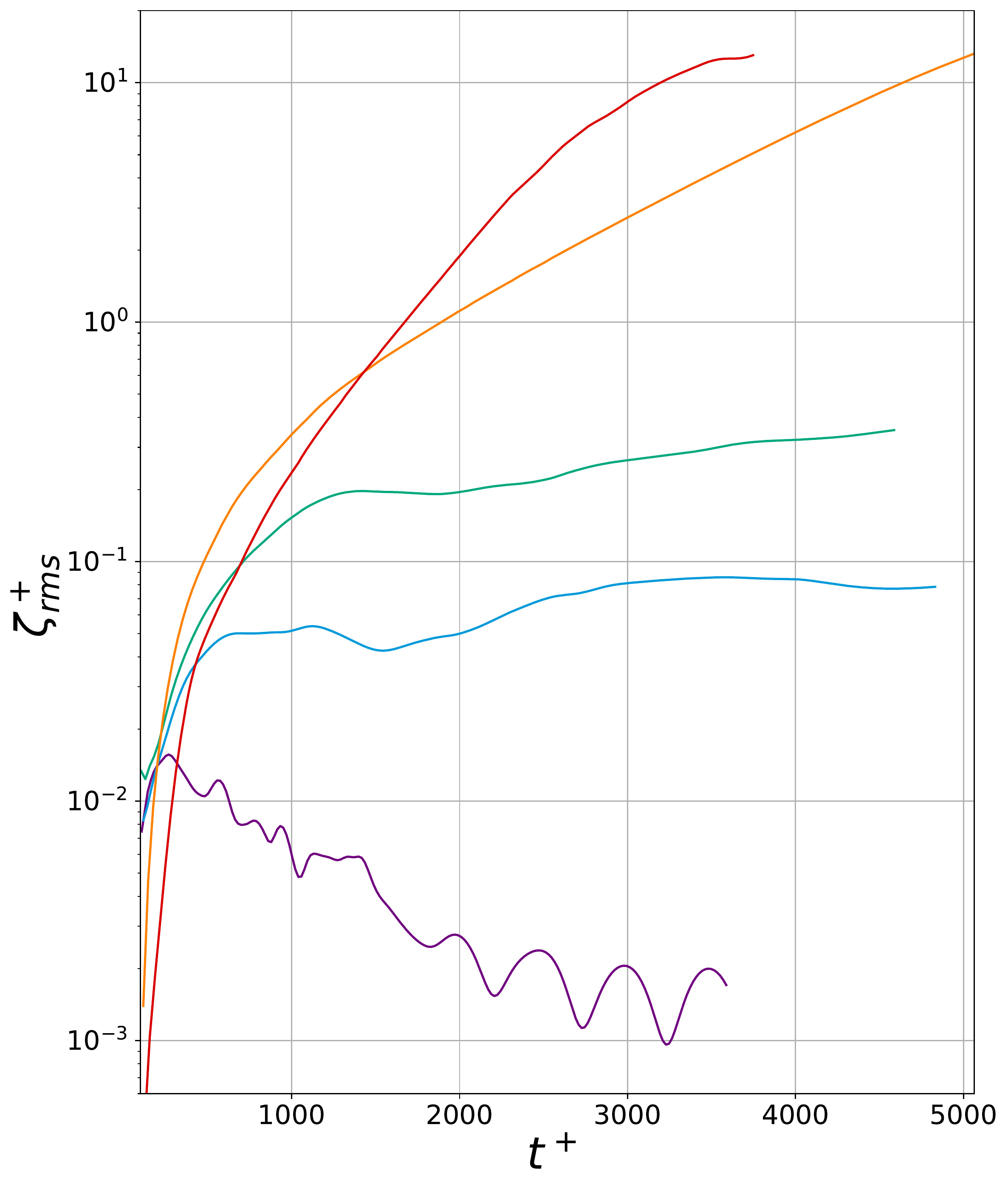}
 \vspace{-1mm}\hspace{5mm}$(a)$\hspace{50mm}$(b)$\hspace{50mm}$(c)$
    \caption{Evolution of the rms interface deformation for different liquid viscosities and wind conditions. $(a)$: water (blue), $\mathcal{L}_{12}$ (green) and $\mathcal{L}_{30}$ (red) under constant wind conditions ($Re^*=800$); $(b)$: water for $Re^*=613$ (purple), $800$ (blue), and $1067$ (cyan); $(c)$: $\mathcal{L}_{100}$ for $Re^*=613$ (purple), $1067$ (blue), $1333$ (green), $1600$ (orange), $1867$ (red). }
      \label{rms}
\end{figure*}
The air flow is driven by the mean pressure gradient $d\overline{\mathcal{P}}/dx$, and the control parameter of the simulations is the friction velocity at the interface, $u^*$, the two being related through the constraint $\rho_au^{*2}=-\frac{H}{2}d\overline{\mathcal{P}}/dx$ resulting from the streamwise momentum balance. A small negative streamwise gravity component, $g_x$, may be introduced in order to impose a zero mean flowrate in the liquid, then mimicking the experiments of \cite{Paquier2016} in which the liquid was enclosed in a rectangular tank. Indeed, the $x$-component of the total driving force per unit volume in the air is $G_a=\rho_ag_x-d\overline{\mathcal{P}}/dx$ while that in the liquid is $G_l=\rho_lg_x-d\overline{\mathcal{P}}/dx$. Since $\rho_l/\rho_a\gg1$, a small but nonzero $g_x$ leaves the driving force virtually unchanged in the air flow ($G_a\approx-d\overline{\mathcal{P}}/dx$) but may provide the dominant contribution in the liquid ($G_l\approx\rho_lg_x$). \\
\indent As long as the flow in the liquid is considered laminar, it is initialized with the Poiseuille profile $\overline{u}_l(z)=\mu_l^{-1}[G_l(z_0-z/2)+\rho_au^{*2}]z$ which satisfies the no-slip condition at the lower wall and the continuity of shear stresses at the mean interface position. In cases a zero net liquid flow rate is assumed, the condition $\int_0^{z_0}\overline{u}_l(z')dz'=0$ yields $G_l=-\frac{3}{2}\rho_au^{*2}z_0^{-1}$. Other prescriptions may be used for $G_l$. For instance, imposing $G_l=-\rho_au^{*2}z_0^{-1}$ yields a zero-shear-stress condition at the bottom of the liquid layer. In the case of water, the subsurface flow is turbulent under most wind conditions. Hence, the above Poiseuille flow and estimates for $G_l$ are physically inadequate. The appropriate initialization procedure is described in \ref{turbliq}.\\
\indent To initialize the air flow, we perform a separate run to compute the single-phase turbulent flow in a plane half-channel of height $H-z_0$, with the friction velocity at the lower wall set to $u^*$. The discretization used in this preliminary run is identical to that described above, i.e. a constant grid spacing $\Delta z_{min}=0.5\nu/u^*$ is employed over a $\Delta/2$-thick region adjacent to the wall, followed by a growing spacing up to the upper free-slip surface.  Once this run has converged, the corresponding velocity and turbulent viscosity fields are pasted in the upper part of the two-phase computational domain, after the velocity of the liquid at the position of the mean interface, $\overline{u}_i=\overline{u}_l(z=z_0)$, has been added to the computed air velocity profile to enforce the continuity of velocities at $z=z_0$. It is worth noting that when the flow in the liquid is laminar and the zero-flowrate condition holds, one has $\overline{u}_i=\mu_l^{-1}[G_lz_0/2+\rho_au^{*2}]z_0=\frac{z_0}{4\mu_l}\rho_au^{*2}$. Then, defining the characteristic Reynolds number in the liquid as $Re_l=\frac{2}{3}\rho_l\overline{u}_iz_0/\mu_l$ (the factor of $2/3$ resulting from the fact that the minimum of $\overline{u}(z)$  stands at the position $z=z_0/3$), $Re_l$ may be shown to be related to the friction Reynolds number of the air flow, $Re^*=\rho_a(H-z_0)u^*/\mu_a$, as $Re_l=\frac{2\rho_l}{3\rho_a}[\frac{z_0}{2(H-z_0)}\frac{\mu_a}{\mu_l}Re^*]^2$. Hence, considering for instance a liquid $10$ times more viscous than water and a configuration with $z_0=H/2$ and $Re^*=800$, one has $Re_l\approx 270$. This estimate indicates that the flow in the liquid remains laminar, even though the air flow above the interface is strongly turbulent; the more viscous the liquid is, the more this conclusion holds.\\
\indent After the above initialization of the velocity and turbulent viscosity fields is completed, the solution of \eqref{divu}-\eqref{SA} starts to be advanced in time. However, to make sure that the matching of velocities and shear stresses at the interface does not introduce any disturbance in the discretized solution, we first let the simulation run from $t=0$ to $t=\Delta t_0=100\nu/u^{*2}$ without updating the volume fraction distribution, $C({\bf{x}},t)$. Then, we add an impulse disturbance, $\zeta_0(x)\delta(t-\Delta t_0)$, to the initial interface position $z_0$, with $\delta(t)$ the Dirac delta function. In practice, the disturbance is applied during one time step. Then we let then the solution of \eqref{divu}-\eqref{SA}, including the volume fraction and local fluid properties, evolve freely for $t>\Delta t_0$. Two specific disturbance shapes have been considered. One is merely the harmonic distribution $\zeta_0(x)=a\cos(kx)$, the wavenumber $k$ being necessarily a multiple of the minimum wavenumber allowed by the length of the domain, $2\pi/L$. To avoid prescribing a specific wavenumber, we generally rather make use of a white noise disturbance defined as
\begin{equation}
\zeta_0(x)=aN^{-1/2}\sum_{n=1}^N\cos(k_nx+2\pi\Phi_n)\,,
\label{wn}
\end{equation}
with $k_n=2\pi n/L$, $k_N=\pi/(2\Delta x)$ and $\Phi_n$ a random phase. This phase is provided by a dedicated routine guaranteeing the same $\Phi_n$-distribution, hence the same disturbance spectrum, in all runs performed on a given grid. An important advantage of \eqref{wn} is that a single run allows us to observe simultaneously the growth of $N_x/2\approx0.6Re^*$ independent wavelengths, as long as effects of nonlinear interactions are negligibly small. For a given $a$, the above two disturbances have identical root-mean square (rms) amplitudes. In both cases, we select $a=2\delta_\nu$, which corresponds to 4 grid cells since $\Delta z_{min}\approx0.5\delta_\nu$. 
Whatever the selected form of $\zeta(x)$, the perturbation of the interface position is translated into a volume fraction disturbance in the form
\begin{equation}
C(x,z,t=\Delta t_0)=\frac{1}{2}\left\{1-\text{erf}\left(\frac{z-z_0-\zeta_0(x)}{\gamma_d\Delta z_{min}}\right)\right\}\,,
\label{C0}
\end{equation}
where $\gamma_d$ is an $\mathcal{O}(1)$-parameter controlling the thickness of the transition region across which the volume fraction changes from $0$ above the interface to $1$ in the liquid ($\gamma_d=1$ in what follows).
\subsection{A few examples of interface evolutions}
\label{evol2}
We finally present some typical results for the evolution of the interface obtained thanks to the approach developed throughout the paper. An extensive discussion of the results and on their consequences with respect to the influence of the liquid viscosity on the growth of surface waves is deferred to a forthcoming publication In what follows, the zero-flow-rate condition is enforced in the liquid in all cases, and the impulse disturbance used to trigger the generation of wind waves obeys the white noise distribution \eqref{wn}. To isolate effects of the liquid viscosity, the other physical properties of the liquid, namely surface tension and density, are kept constant and equal to values corresponding to water. For the same reason, the flow in the liquid is considered laminar in all runs, although in the case of water it would actually be turbulent under most wind conditions. In what follows, a liquid $q$ times more viscous than water will be denoted as $\mathcal{L}_q$ for the sake of conciseness. Surface deformations and times are normalized by the viscous near-wall scales of the air flow, $\delta_\nu=\nu/u^*$ and $t_\nu=\nu/u^{*2}$, respectively, and the corresponding normalized quantities are denoted with a $+$ superscript.\\
\begin{figure*}[t!]
    \center
    \hspace{-7mm} \includegraphics[width=0.34\textwidth]{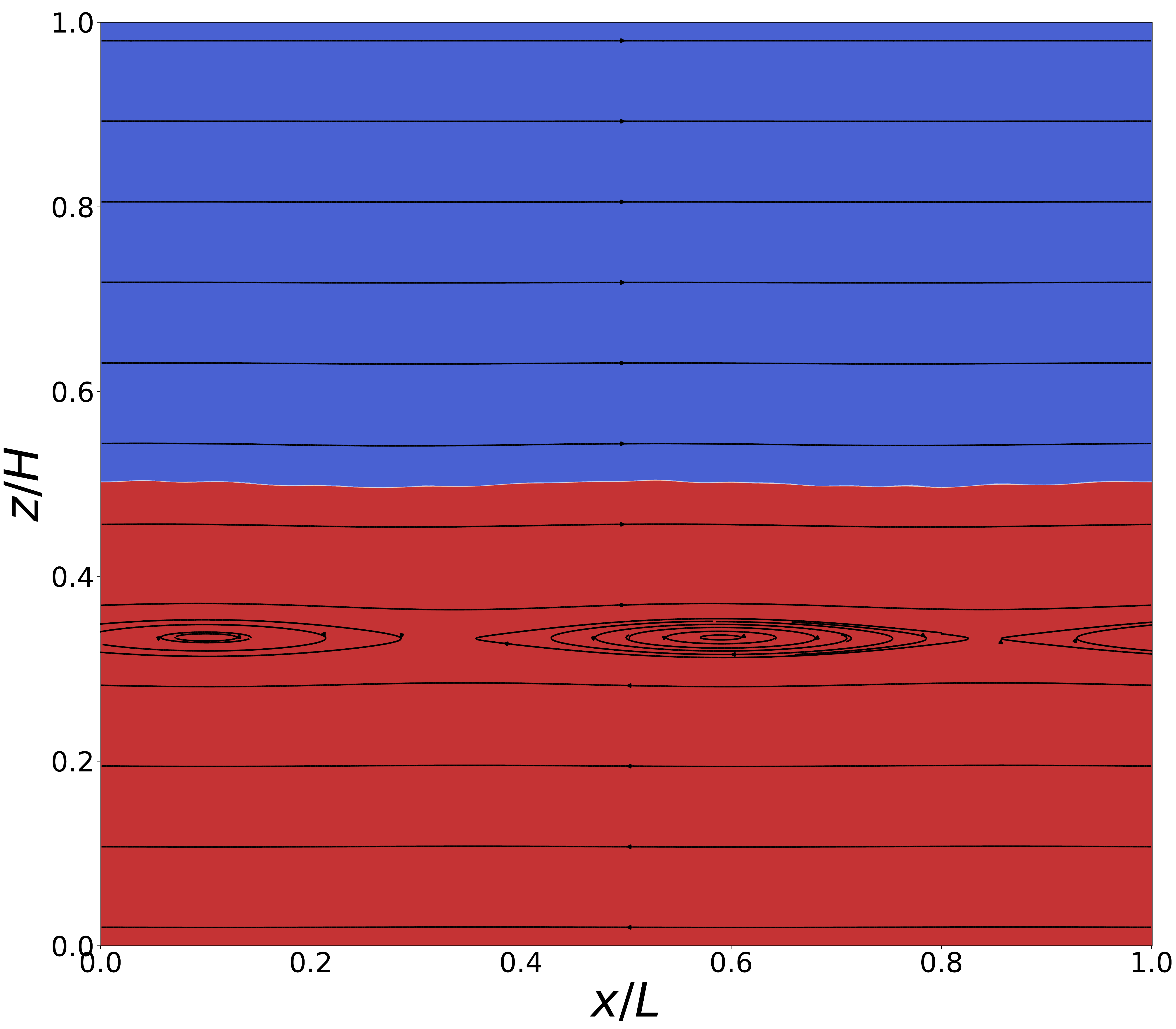}
     \hspace{-1mm} \includegraphics[width=0.34\textwidth]{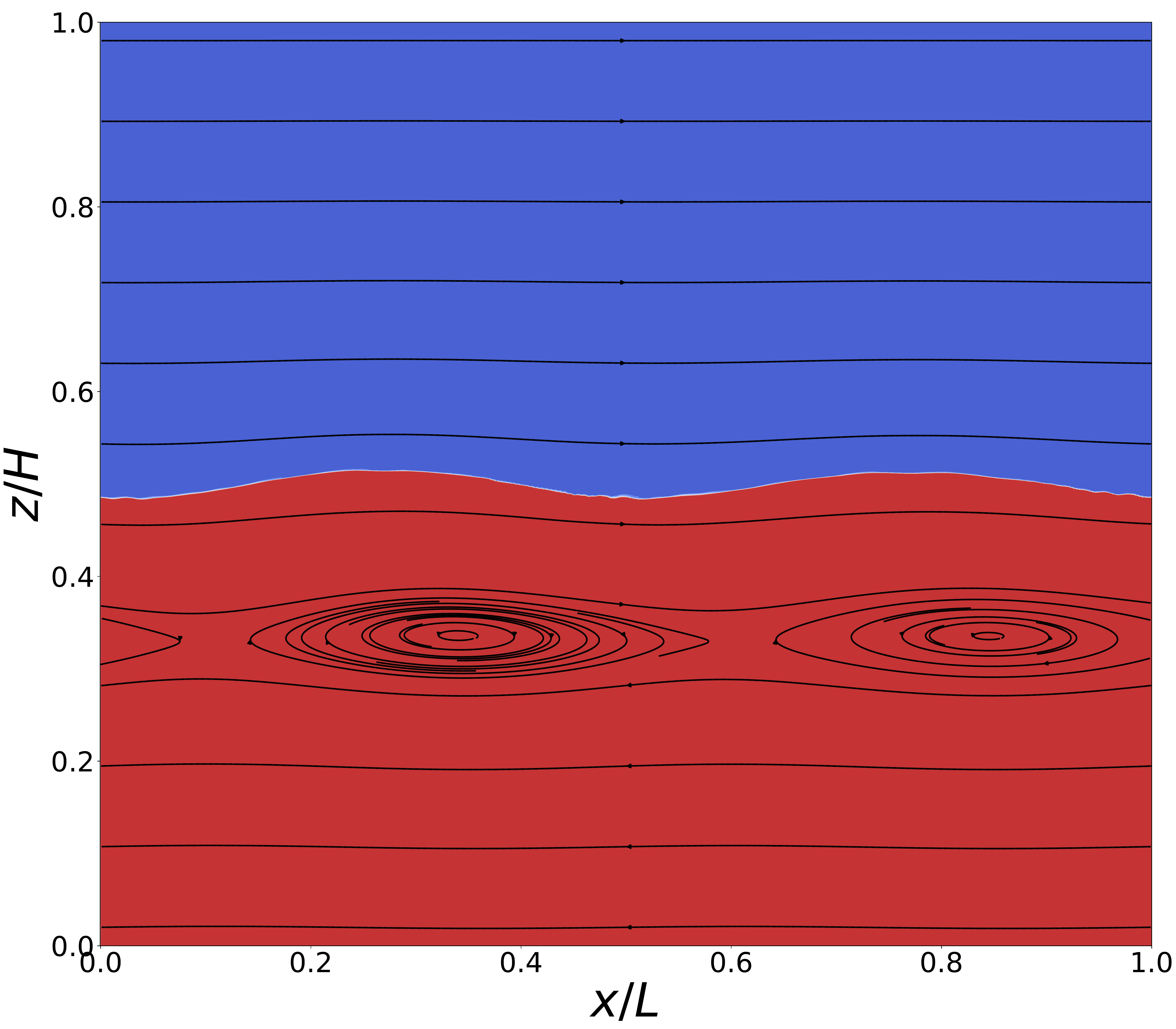}
   %    \vspace{-1mm}\hspace{10mm}$(a)$
     \hspace{-0.5mm}\includegraphics[width=0.34\textwidth]{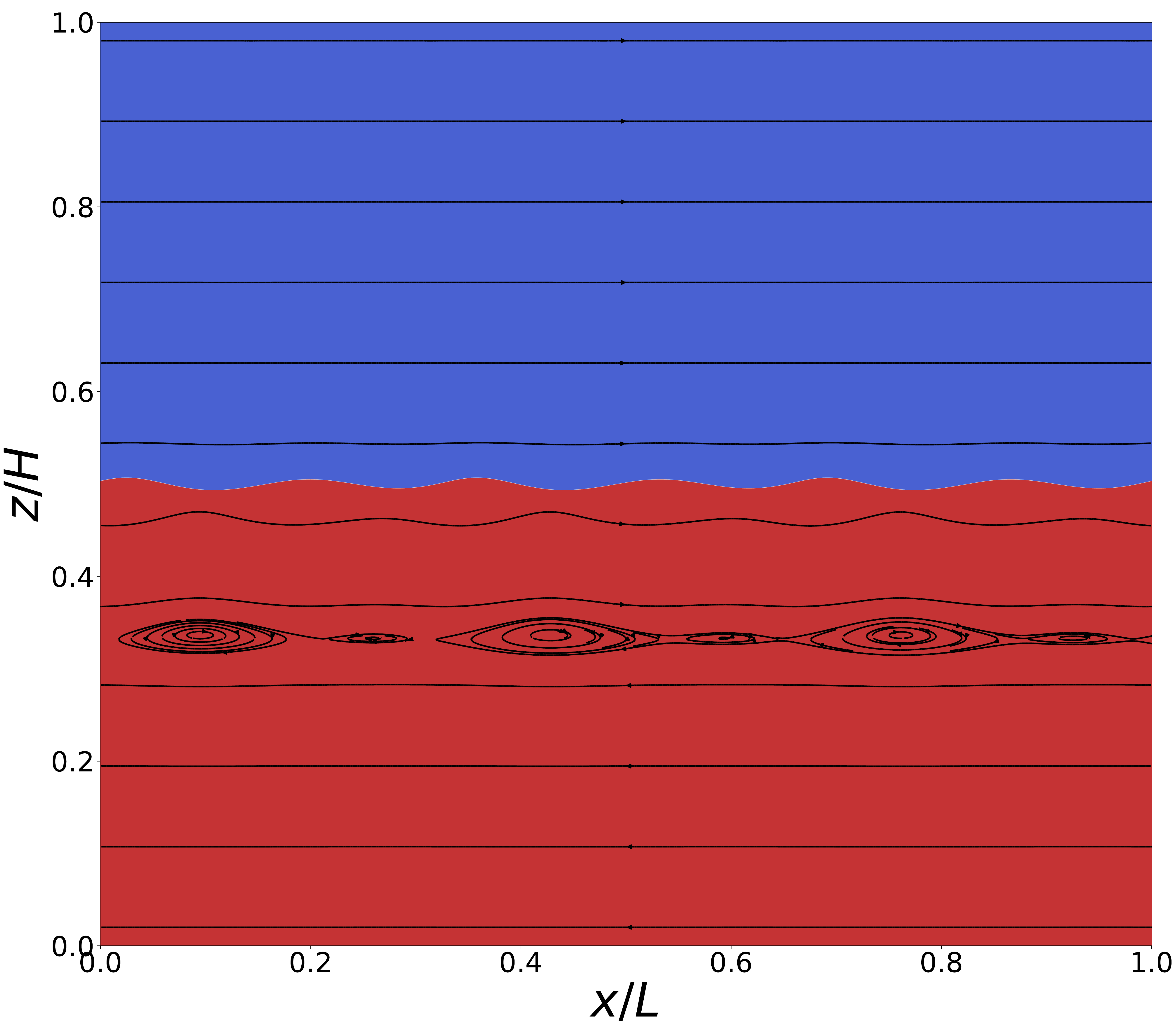}\\
  %    \vspace{-1mm}\hspace{10mm}$(b)$
 \vspace{0mm}\hspace{-0mm}$(a)$\hspace{50mm}$(b)$\hspace{50mm}$(c)$
    \caption{Snapshots of the interface deformation and streamline patterns in the laboratory frame. $(a)$: water for $Re^*=800$ at $t^+=6000$; $(b)$: same at $t^+=7000$; $(c)$: $\mathcal{L}_{100}$ for $Re^*=1600$ at $t^+\approx5000$. The liquid (red) and air (blue) layers are defined as the regions where the instantaneous volume fraction $C({\bf{x}},t)$ is larger or smaller than $0.5$, respectively.}
      \label{stream}
\end{figure*}
\indent Figure \ref{rms}$(a)$ shows how the same initial interface disturbance submitted to the same wind speed $(Re^*=800)$ evolves for three liquids of increasing viscosity, namely water, $\mathcal{L}_{12}$ and $\mathcal{L}_{30}$. With water, two stages during which $\zeta_{rms}^+$ exhibits a sharp, nearly exponential increase emerge, separated by an intermediate stage corresponding to a milder increase from $t^+\approx3000$ to $t^+\approx5000$. As we shall see later, the two exponential growths are associated with disturbance components having distinct wavenumbers. At the end of the simulation ($t^+\approx8000$), the surface deformation is still growing exponentially and its rms value has already reached $50$ viscous length units. In the case of $\mathcal{L}_{12}$, the interface evolution follows a different route. Here also, three distinct stages may be identified but the rms deformation now exhibits its strongest growth rate during the short intermediate period. The deformation is still growing at $t^+=6000$ but the growth rate is decreasing and this trend persists at later stages ($\zeta_{rms}^+\approx0.5$ at $t^+=12000$). Therefore, the saturated wave amplitude is expected to be of the order of the viscous near-wall length scale in that case. Last, after having slightly grown up to $t^+\approx3000$, the deformations at the surface of $\mathcal{L}_{30}$ start to decay. This trend goes on throughout the simulation and the rms deformation at $t^+=12000$ is typically one order of magnitude weaker than the broad maximum reached at $t^+\approx3000$. Consequently, with $Re^*=800$, present simulations predict that large-amplitude waves develop at the surface of water, while in the case of $\mathcal{L}_{12}$ waves barely reach an amplitude of the order of $\delta_\nu$, and disturbances at the surface of $\mathcal{L}_{30}$ first grow transiently but then experience a continuous decay.\\
\indent Figure \ref{rms}$(b)$ shows how the deformation varies with the wind speed at the surface of water in a range of conditions corresponding approximately to dimensional bulk air velocities from $4.5\,$m.s$^{-1}$ to $8\,$m.s$^{-1}$, the latter being the highest speed considered in \cite{Paquier2016}. Not surprisingly, the higher the wind speed the stronger the growth rate, i.e. the larger $\zeta_{rms}^+$ at a given $t^+$. One may also notice that the three evolutions exhibit common trends, especially the three-stage structure already noticed in figure \ref{rms}$(a)$, with an intermediate stage during which the growth rate is significantly weaker than at shorter and longer times. Actually, the growth rate may even vanish during that stage at low enough wind speed, as the curve corresponding to $Re^*=613$ indicates. The case of $\mathcal{L}_{100}$ is considered in figure \ref{rms}$(c)$. After the initial transient, the rms deformation quickly decays at the lowest wind speed ($Re^*=613$), while it stays almost constant and of the order of one-tenth of $\delta_\nu$ for $Re^*=1067$ (note that this is the wind condition yielding the largest growth for water in figure \ref{rms}$(b)$). Increasing the wind speed, the interface deformation is found to exhibit a more consistent growth for $Re^*=1333$, but its rms value is still significantly less than $\delta_\nu$ at the end of the simulation. In contrast, the highest two wind conditions ($Re^*=1600$ and $Re^*=1867$) give rise to a continuous growth, with a growth rate decreasing gradually over time and  yielding amplitudes larger than $10\delta_\nu$ at the end of the computation. It is worth noting that the three-stage structure observed with water does not exist in the case of $\mathcal{L}_{100}$, which suggests qualitative differences in the evolution of the interface deformation in the presence of low- and high-viscosity liquids. \\
\begin{figure*}[t!]
    \center
    \hspace{-7mm} \includegraphics[width=0.34\textwidth]{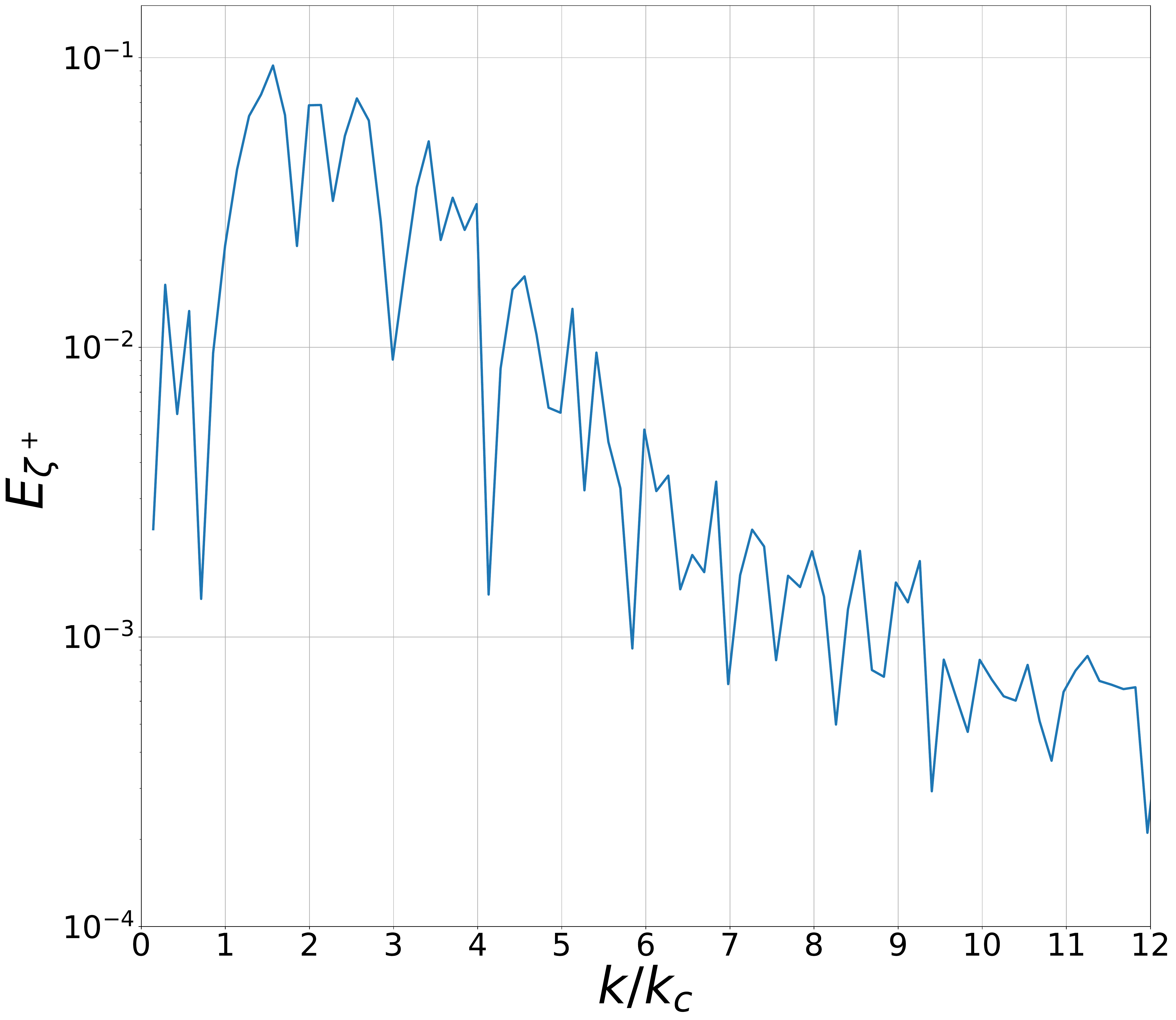}
     \hspace{-1mm} \includegraphics[width=0.34\textwidth]{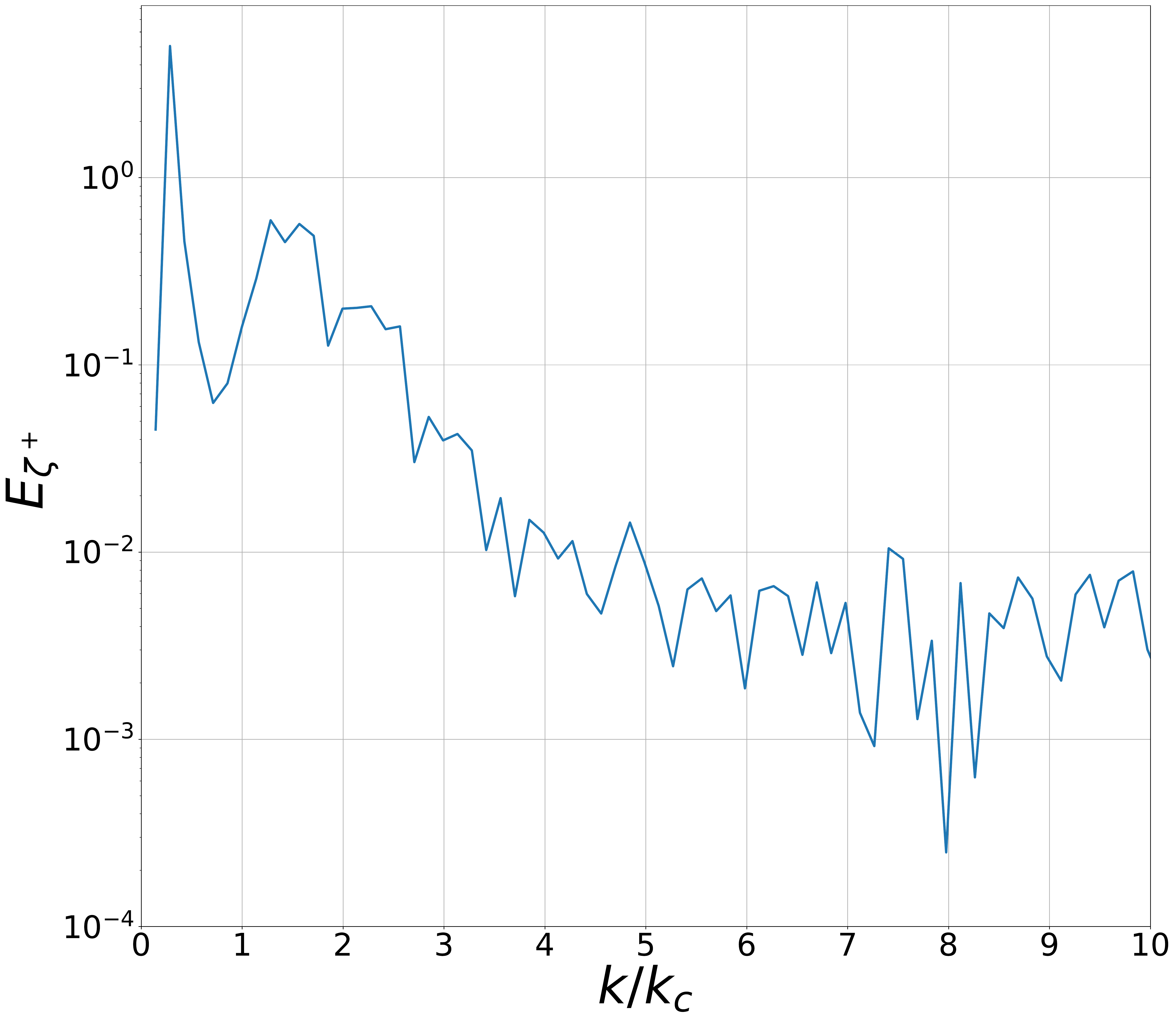}
   %    \vspace{-1mm}\hspace{10mm}$(a)$
     \hspace{-0.5mm}\includegraphics[width=0.34\textwidth]{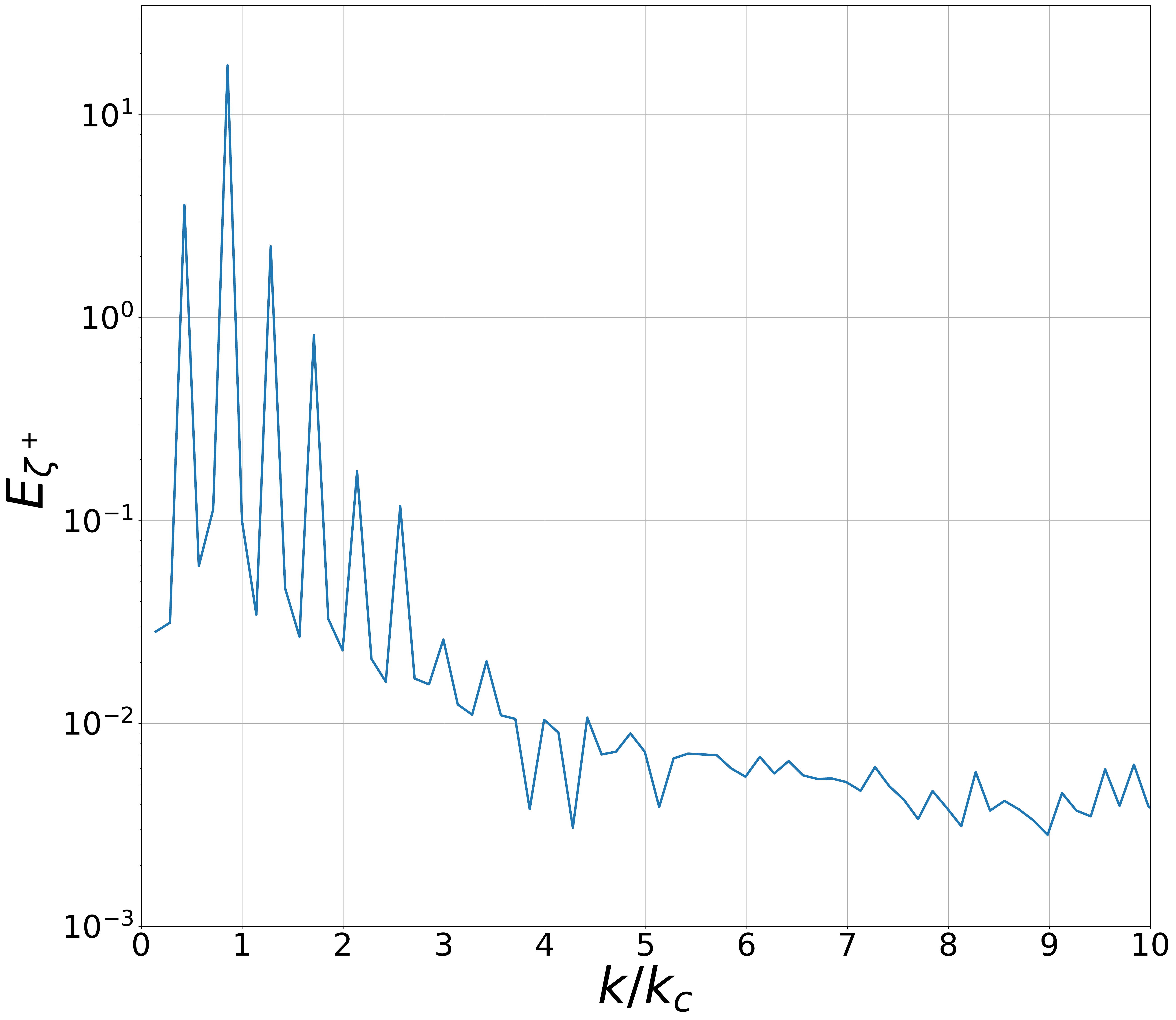}\\
  %    \vspace{-1mm}\hspace{10mm}$(b)$
 \vspace{0mm}\hspace{-0mm}$(a)$\hspace{50mm}$(b)$\hspace{50mm}$(c)$
    \caption{Power spectral density $E_{\zeta^+}(k/k_c)$ of the wave field. $(a)$: water for $Re^*=800$ at $t^+=2000$; $(b)$ same at at $t^+=6000$; $(c)$: $\mathcal{L}_{100}$ for $Re^*=1600$ at $t^+=5000$. Wavenumbers are normalized by the capillary wavenumber $k_c=(\rho_lg/\gamma)^{1/2}$.}
      \label{spec}
\end{figure*}
\indent Some examples of the interface and streamline patterns after the interface deformation has reached a sizable amplitude are displayed in figure \ref{stream}. The wavelength $\lambda=L/2$ is seen to dominate the wave field at such long times in the case of water (figures \ref{stream}$(a-b)$), while a shorter dominant wave with $\lambda=L/6$ has emerged in the case of $\mathcal{L}_{100}$ (figure \ref{stream}$(c)$). The corresponding wave steepness is close to $0.2$ and $0.275$, respectively, indicating that waves have already reached a stage in which nonlinear effects are strong. A series of closed streamlines is seen to exist in the liquid in all cases, and stands approximately two thirds of the height of the liquid. This feature is due to the zero-flowrate condition that forces the bottom part of the liquid to recirculate. The vertical positions of the centroids of the closed streamlines almost coincide with those of the wave troughs in the case of $\mathcal{L}_{100}$, while they are slightly ahead of the positions of the crests in the case of water. Similarly, the crests and troughs of the first streamline beneath the interface nearly coincide with those of the latter in the case of water, while the undulations of this streamline and those of the interface are almost phase-opposed for $\mathcal{L}_{100}$. The reason for this difference may be qualitatively understood by noting that, owing to the presence of the $\mu_l^{-1}$ pre-factor in the expression of the surface mean velocity $\overline{u}_i$,  the phase speed $c_0$ of the dominant wave with wavenumber $k_0$, roughly estimated as $c_0=(g/k_0)^{1/2}$, is approximately five times larger than $\overline{u}_i$ in figure \ref{stream}$(c)$, while it is only one-third of it in figures \ref{stream}$(a-b)$. Therefore, a critical layer exists in the latter case while it does not in the former. Consequently, waves propagating at the surface of $\mathcal{L}_{100}$ are nearly irrotational, while those propagating over the water layer have a significant vortical component. The stream function of an irrotational monochromatic wave at a given $z$ is known to reach its maxima below the crests and its minima below the troughs, which, when observed in the fixed laboratory frame, translates into the phase-opposed pattern displayed in figure \ref{stream}$(c)$. Conversely, since the ratio $c_0/\overline{u}_i$ is small in the case of water, the corresponding interface deforms slowly in the laboratory frame and is then close to a streamline. This is why the streamline distribution in figures \ref{stream}$(a-b)$ is reminiscent of the Kelvin's cat's eye pattern. Obviously, the streamline pattern observed in figures \ref{stream}$(a-b)$ would be significantly modified if the flow in the water were treated as turbulent, as it is in practice at the considered wind speed. In that case, the approximate model developed in \ref{turbliq} indicates that the surface current would be approximately 6 times weaker than predicted assuming laminar conditions, so that no critical layer would exist in the liquid. Influence of the laminar vs turbulent nature of the wind-driven subsurface current in the case of water will be discussed in a forthcoming paper.\\
\indent Figure \ref{spec} shows some typical wave energy spectra, $E_{\zeta^+}(k/k_c)$, still for the two cases of water at $Re^*=800$ and $\mathcal{L}_{100}$ at $Re^*=1600$. The capillary wavenumber $k_c=(\rho_lg/\gamma)^{1/2}$, kept constant in all runs, is used to normalize wavenumbers. In the fairly early stage displayed in figure \ref{spec}$(a)$, several wavenumbers located in the range $1.2\leq k/k_c\leq2.8$ dominate the wave field, the energy peak being reached at $k_0/k_c\approx1.55$, i.e. for a wavelength $\lambda_0\approx0.09L$. The energy contained in lower wavenumbers such that $k/k_c<1$ is typically one order of magnitude weaker. Beyond $k/k_c\approx3$, the energy density decays gradually as the wavenumber increases, until $k/k_c\approx10$ where it is two orders of magnitude weaker than the peak value. Then it remains at similar levels down to the highest wavenumber allowed by the grid (not shown). As time proceeds, the spectral content in the range $k/k_c\lesssim10$ changes dramatically. In subfigure $(b)$, i.e. at a time instant thrice as long as in subfigure $(a)$, a single component dominates the wave field, carrying an energy density one order of magnitude larger than that found in the previously dominant wavenumber range. This component corresponds to $k_0/k_c\approx0.285$, i.e. $\lambda_0=L/2$. Therefore, the dominant wavenumber has been reduced by a factor of $5.55$ in between the two snapshots. This wavenumber downshift is the counterpart of the well-known frequency downshift observed with spatially growing waves, especially at short fetch \citep{Lake1977,Su1982,Huang1996}. Frequency downshift is known to be a consequence of the Benjamin-Feir instability. Dissipative effects such as wave breaking and generation of capillary ripples were readily identified as the possible relevant sources of dissipation in mechanically-generated waves \citep{Lake1977,Melville1982}. Nevertheless, purely conservative three-dimensional effects were later shown to strongly promote the phenomenon \citep{Trulsen1997}, although the possibility that they may produce the observed downshifts without the presence of any source of dissipation was debated \citep{Trulsen1997,Dias1999}. With wind-generated waves, turbulence in the air and possibly in the liquid adds another obvious source of dissipation. Assuming a logarithmic velocity profile in the air and a constant eddy viscosity in both air and water, it has been shown that frequency downshifts in reasonable agreement with experimental data may be predicted in two-dimensional configurations involving both pure gravity waves \citep{Hara1991} and gravity-capillary waves \citep{Hara1994}. Here, three-dimensional effects are absent and wave breaking does not occur in the stages corresponding to figures \ref{spec}$(a-b)$. Moreover, the flow in the liquid is artificially forced to stay laminar and the corresponding viscous dissipation in the bulk is negligibly small given the low viscosity of water. Therefore the only significant source of dissipation stands in the turbulence in the air boundary layer, possibly supplemented with some viscous near-surface dissipation in water due to the presence of steep high-wavenumber capillary ripples in the trough region of the carrying gravity wave (see figure \ref{stream}$(b)$). What present results show is that large downshifts take place even within this restrictive modelling framework. Comparing these predictions with laboratory observations showing how the dominant wave length grows with increasing fetch, for both water and more viscous liquids, is of course an important test for the present approach, including the selected one-equation turbulence model.\\
\indent Figure \ref{spec}$(c)$ reveals a very different spectral distribution of the wave energy in the case of $\mathcal{L}_{100}$. Here, most of this energy is contained in a discrete series of rays consisting of a dominant component $k_0/k_c\approx0.855$, i.e. $\lambda_0=L/6$, its subharmonic $k_0/2$, and the successive harmonics $3k_0/2$, $2k_0$, etc.  Energy has decayed by three orders of magnitude in between the dominant component and the eighth spectral ray, corresponding to the fourth harmonic $k=4k_0$.
The energy distribution found in figure \ref{spec}$(c)$ is a clear indication that intense subharmonic \cite{Longuet-Higgins_1978b,Su1982} and superharmonic \cite{Longuet-Higgins_1978a} instabilities may affect wind-generated waves propagating over high-viscosity liquids, as is customary with nonlinear irrotational surface waves. Since the wave steepness is close to $0.3$ here, the presence of such strong nonlinear effects is actually no surprise. 
%the curvature of the mean velocity profile in the liquid is $-G_l/\mu_l=\frac{3}{2}\rho_gu^{*2}(z_0\mu_l)^{-1}$, a quantity much larger for water than for $\mathcal{L}_{100}$, even though the friction velocity is twice as large in the latter case.  In contrast, in the case of water, the influence of the curvature of this current cannot be neglected. The stream function associated with the interface disturbance then obeys the classical Rayleigh equation of hydrodynamic stability theory and the corresponding velocity field is no longer irrotational. Under such conditions, the stream function 

\section{Summary and concluding remarks}
\label{conclu}
In this paper, we reported the successive steps of the development of a consistent numerical approach aimed at computing the evolution of the interface separating a viscous liquid layer from a turbulent air flow blowing over it. The final aim was to obtain a numerical set up capable of providing predictions for the growth of wind-generated waves developing at the surface of a liquid which may be several orders of magnitude more viscous than water, which implies high bulk velocities in the air, hence a broad spectrum of eddies within the boundary layer. After reviewing the strengths and limitations of the various available simulation approaches, these flow conditions led us to favour the combination of a one-fluid formulation of the entire time-evolving flow field, combined with a Reynolds-averaged formulation of the governing equations. Given that the time scales of the orbital and turbulent motions generally largely overlap, we showed that the simplest way to combine these two formulations in a consistent manner is achieved by defining the averaging operator as a spatial average in the transverse (spanwise) horizontal direction. We discussed the restrictions implied by the inherently two-dimensional approach resulting from this choice, especially the fact that it is unsuitable for studying the occurence and development of tiny three-dimensional wrinkles at the interface, and to examine the relevance of Phillips resonance mechanism in the very early stages of the interface deformation. In contrast, Miles instability mechanism being primarily two-dimensional, the designed modelling framework is appropriate for examining its relevance over a wide range of liquid viscosities and air flow conditions.\\
\indent Due to the selected Reynolds-averaged formulation, turbulent stresses appear in the resulting momentum equations. The potentialities of the proposed approach then depend to a significant extent on the choice of an appropriate turbulence closure for these stresses, and in a second step of the selected turbulence model. Although deficiencies of eddy-viscosity type closures in the outer part of the boundary layer above a wavy wall have been identified for a long time, we adopted such a closure here and selected the one-equation Spalart-Allmaras model to compute the eddy viscosity throughout the flow. Besides its simplicity and numerical robustness, this model has the advantage that the transported quantity, i.e. the eddy viscosity, vanishes at the interface since turbulent fluctuations in the air flow `feel' it as a rigid wall, given the very large liquid-to-air density ratio. %The same conclusion regarding the eddy viscosity holds on the liquid side if the subsurface current is turbulent, although the tangential velocity fluctuations do not vanish. 
More sophisticated alternatives, i.e. two-equation and full Reynolds-stress models, all solve a transport equation for a lengthscale-determining quantity, such as the dissipation rate, which exhibits a more complex near-wall behaviour, a feature complicating the description of the viscous sublayer which is of special importance in the present problem.  \\
\indent The reported tests of the turbulence model in single-phase wall-bounded flows revealed its strengths and weaknesses. Among the former, the model was shown to produce accurate near-wall mean velocity distributions in a fully-developed channel flow, even with a sparse number of cells in the viscous and buffer layers. In the range of wavenumbers relevant to the generation of short (i.e. young) wind waves, it also correctly predicts the phase shift of the wall pressure and shear stress distributions in the unseparated flow over a wavy wall. Similarly, it predicts quite well the location of the detachment and reattachment points in the separated flow over a large-amplitude wavy wall. The main shortcoming of the model in this type of flow appears to be its significant under-estimate of the maxima of the wall pressure and shear stress, especially that of the excess surface shear in the windward region ahead of the wave crest. Obviously, one may expect this shortcoming to have some consequences on the predicted growth rate of the interface deformations, and this is an issue to be examined in detail in the next steps of this study.\\
\indent Having qualified the turbulence model in single-phase flows, we turned to two-phase configurations. In the examples we considered, the flow is assumed to be turbulent above the interface and laminar in the liquid, obeying a zero-flowrate condition in the latter. We paid special attention to the initialization procedure, to make sure that the initial composite velocity field is a stationary solution of the discretized two-phase problem as long as the interface remains flat. Then, we introduced a small impulse disturbance in the interface position, the spectral content of which corresponds either to a monochromatic wave or to a white noise distribution. We then presented a short selection of results based on evolutions of the rms disturbance, streamline and interface patterns, and energy spectra of the wave field for liquids of various viscosities. These results revealed the key influence of the liquid viscosity on the nature and structure of the observed evolutions. Obviously, for a given air friction velocity, the higher the liquid viscosity the lower the growth rate. Conversely, for a given liquid, the higher the friction velocity the stronger the growth rate. Provided the air flow is strong enough to make waves develop at the surface of a liquid one hundred times more viscous than water, we found that the corresponding growth rate decreases continuously over time and the wave field is essentially composed of a series of rays dominated by a carrier wave, the subharmonic with a wavenumber of half, and some higher harmonics resulting from the combination of these two and from nonlinear interactions. In contrast, in the case of water, a strong wavenumber downshift takes place as the disturbance grows, resulting in two separate stages during which the corresponding dominant ray grows exponentially. This downshift, together with the growth rate and critical wind speed below which the disturbance is damped, are among the main indicators to be compared with experimental measurements in a forthcoming paper. The influence of the subsurface current, which was artificially kept laminar in the presented runs performed with water, but would actually be turbulent under natural conditions, also deserves specific attention. Indeed, we observed that the large surface velocity resulting from the laminar assumption induces the presence of a critical layer beneath the interface, while such a feature does not exist if the subsurface current is turbulent. This difference may have consequences on the wave growth and even on the selected preferential wavenumber.\\
\indent Improvements in the numerical approach presented here will presumably mostly come from the introduction of a more accurate turbulence model. However, which model offers the best predictability with respect to the targeted problem while guaranteeing a good numerical stability is still to be determined. In particular, it seems that preserving the efficiency of the one-equation model close to the interface, which is due to the combination of the physical ingredients it incorporates and of the versatile technique we implemented to compute the local distance from an arbitrarily-shaped interface, would make sense. However, a Reynolds-stress transport model is clearly required further away from the interface to properly capture the out-of-equilibrium effects resulting from the distortion imposed by the growing waves on the large-scale turbulent eddies. A zonal approach in which these two types of models would be matched at some position lying in the logarithmic layer might offer the best compromise. Nevertheless, detailed tests are still necessary to reach a firm conclusion on the best turbulence modelling strategy.
%As we imposed a zero-flowrate condition in the liquid, the flow below the waves is laminar, yielding a significant a critical layer was found to exist below waves propagating at the surface of water but not at that of a hundred times more viscous liquid, yielding 

\subsection*{CRediT authorship contribution statement}
%\textcolor{red}{A affiner en fonction de la nomenclature impos\'ee par le journal :}\\
\textcolor{black}{{\bf{Florent Burdairon:}} Investigation, Validation, Formal analysis, Writing - original draft.
{\bf{Jacques Magnaudet:}} Conceptualization, Formal analysis, Writing - review \& editing, Supervision, Funding acquisition.}

\section*{Declaration of competing interest}
\textcolor{black}{The authors declare that they have no known competing financial interests or personal relationships that could have appeared to influence the work reported in this paper.}
\section*{Acknowledgements}
\textcolor{black}{The specific computational developments described in the paper owe much to the continuous support of P. Elyakime. F. Burdairon's fellowship was
granted by the ViscousWindWaves project of the French National Research Agency (Project No. ANR-18-CE30-0003). Computations were made possible by the allocation of HPC resources in the CALMIP supercomputing meso-center under grant P22032.}

\appendix

\section{Detail of the turbulence model}
\label{SAM}
In \eqref{SA}, the wall-corrected vorticity magnitude $ \tilde{\Omega}$ involved in the production term is defined as \citep{SA2012}
 \begin{eqnarray}
 \tilde{\Omega}&=&\Omega+\Omega_\ell\quad\mbox{if}\quad \Omega_\ell/\Omega>-c_{v2}\,,\\
 \nonumber
  \tilde{\Omega}&=&\Omega\Bigg\{1+\frac{c_{v2}^2\Omega+c_{v3}\Omega_\ell}{(c_{v3}-2c_{v2})\Omega-\Omega_\ell}\Bigg\}\quad\mbox{otherwise,}
 \end{eqnarray}
 with the near-wall correction $\Omega_\ell$ such that
\begin{equation}
\Omega_\ell=f_{v2}\frac{\tnu}{\kappa^2\ell^2}\,.
\label{omegatilde}
\end{equation}

The damping functions $f_{v1}$ (involved in \eqref{nutilde} and \eqref{SAD}) and $f_{v2}$ are prescribed as
\begin{equation}
     f_{v1} = \frac{(\tnu/\nu)^3}{(\tnu/\nu)^3 + c_{v1}^3}\quad\mbox{and}\quad   f_{v2} = \frac{1+(f_{v1}-1)(\tnu/\nu)}{1+f_{v1}(\tnu/\nu)} \,,   % \quad       \chi = \frac{\tnu}{\nu}
    \label{Cornu}
\end{equation}
so that $ f_{v1} \rightarrow1$ and  $f_{v2}\rightarrow0$ for large $\tnu/\nu$, and $ f_{v1} \rightarrow0$ and  $f_{v2}\rightarrow1$ when $\tnu/\nu\rightarrow0$. Similarly, the function $f_w$ in the wall-destruction term of (\ref{SA}) is requested to become unity at the wall and to vanish far from it. It is defined as
\begin{eqnarray}
f_w&=&g\Bigg(\frac{1+c_{w3}^6}{g^6+c_{w3}^6}\Bigg)^{1/6}\,,\\
\nonumber
\mbox{with}\,\, g&=&r[1+c_{w2}(r^5-1)]\quad\mbox{and}\,\, r=\frac{\tnu}{\kappa^2\tilde{\Omega}\ell^2}\,.
\end{eqnarray}
Last, the prescribed values for the empirical constants involved in the model are 
\begin{eqnarray}
\label{const}
\nonumber
c_{b1}&=&0.1355\,,\, c_{b2}=0.622\,,\,  \sigma=2/3\,,\\
c_{v1}&=&7.1\,,\,c_{v2}=0.7\,,\,c_{v3}=0.9\,,\\
\nonumber
c_{w1}&=& \frac{c_{b1}}{\kappa^2}+\frac{1+c_{b2}}{\sigma} \approx3.24\,,\\
\nonumber
c_{w2}&=&0.3\,,\, c_{w3}=2.0\,,
\end{eqnarray}
and the von K\'{a}rm\'{a}n constant $\kappa$ is set to $0.41$. \\
\indent In the original version of the model, the function $f_r$ weighting the production term in \eqref{SA} is unity. Since then, the model was checked in flows involving curved streamlines \citep{Spalart1997,Shur2000} and it was suggested that its performance may be improved by making $f_r$ depend on the relative magnitude of rotation and strain effects. %The corresponding formulation involves the Lagrangian derivative of the strain-rate tensor, leading to a complex implementation and to quite high computational costs. 
A simple formulation for such a correction, involving the norm of the rotation- and strain-rate tensors, was proposed in \cite{Hellsten1998} in the context of a two-equation turbulence model. The idea is to define a curvature/rotation Richardson number $\frac{\Omega}{\mathcal{S}}\Big(\frac{\Omega}{\mathcal{S}}-1\Big)$ comparing the time scales associated with the rotational and straining contributions to the local velocity gradient, $\mathcal{S}=(2\,\langle{\bf{S}}\rangle\colon\langle{\bf{S}}\rangle)^{1/2}$ denoting the norm of the strain-rate tensor. This formulation was adapted to the Spalart-Allmaras model in \cite{Qiang2013}, %and was shown to be as successful as the one originally proposed in \cite{Spalart1997} in two-dimensional geometries with highly curved walls. In this approach
with the weighting function defined as
\begin{equation}
\label{fr}
f_r=\big(1+c_{r1}\big)\frac{2\mathcal{S}}{\mathcal{S}+\Omega}\Bigg\{1-c_{r3}\tan^{-1}\Bigg[c_{r2}\frac{\Omega}{\mathcal{S}}\Bigg(\frac{\Omega}{\mathcal{S}}-1\Bigg)\Bigg]\Bigg\}-c_{r1}
\end{equation}
with %$\text{S}=(2\,{\bf{S}}\colon{\bf{S}})^{1/2}$ the norm of the strain-rate tensor and
\begin{equation}
c_{r1}=1.0\,,\,c_{r2}=2.0\,,\,c_{r3}=1.0\,.
\label{constr}
\end{equation}
In thin shear flows, $\Omega=\mathcal{S}$, so that $f_r=1$, which leaves the magnitude of the production term in (\ref{SA}) unchanged. In contrast, $f_r$ is larger than unity in strain-dominated regions (e.g. $f_r(\Omega/\mathcal{S}=0.8)\approx1.91$), increasing the magnitude of the production term, hence that of $\nu_t$. Conversely, $f_r$ is smaller than unity in vorticity-dominated regions and becomes negative beyond $\Omega/\mathcal{S}\approx1.2$, extinguishing turbulence in vortex cores. Since $\Omega/\mathcal{S}$ is unity at the wall, so is $f_r$. Hence, in (\ref{const}), the relation linking $c_{w1}$ to $c_{b1}$, $c_{b2}$, $\sigma$ and $\kappa$ remains unchanged.
 % In \cite{SA2012} it was pointed out that $\Omega_\ell$, hence $\Omega$, may become negative in flow regions corresponding to moderate turbulence levels and low vorticity. To avoid such an unphysical behaviour, it was proposed to apply (\ref{tomega}) in regions where $\Omega_\ell/\Omega>-c_{v2}$, and to 
%\begin{equation}
 %   P = c_{b1} \widetilde{S} \tnu       \qquad      D = c_{w1} f_w \left[ \frac{\tnu}{d} \right]^2
%    \label{eq:spalart_allmaras_termes}
%\end{equation}

\section{Approximate characteristics of a turbulent liquid layer beneath a sheared interface}
\label{turbliq}
Since the viscosity ratio $\mu_l/\mu_a$ is always large, even with water, turbulence beneath an air-liquid interface subjected to a prescribed shear stress does not obey a no-slip condition, unless the surface is contaminated by surfactants. For this reason, nonzero tangential velocity fluctuations subsist in the liquid all the way to the interface. In contrast, the normal fluctuation vanishes on it, provided the turbulence intensity  is not strong enough to distort the surface. These combined behaviours yield a linear growth of the turbulent shear stress as the distance to the interface, $z_0-z$, increases. This difference with the more familiar situation of a no-slip wall has direct consequences on the structure of the near-surface flow. In particular, compared with the boundary layer adjacent to a no-slip wall, the viscous sublayer is almost twice as thin and the buffer layer is virtually absent \citep{McLeish1975,Tsai2005,Enstad2006}, so that the log-law for the mean velocity profile applies much closer to the surface. This is why setting the damping function $f_{v1}$ involved in \eqref{SAm} and \eqref{Cornu} to unity in the liquid is sufficient to obtain a realistic description of the near-interface region with the Spalart-Allmaras model. This is also why a simple approximate eddy-viscosity profile valid throughout the liquid layer can be derived. \\
\indent For this, let us first consider that the subsurface flow consists in a layer with kinematic viscosity $\nu_l=\mu_l/\rho_l$ extending from the interface $z=z_0$ at which the shear stress $\rho_l u_l^{*2}\equiv\rho_a u^{*2}$ applies, down to a lower plane $z=\delta_lz_0$ at which the shear stress vanishes (the index $l$ is used to denote quantities in the liquid in cases an ambiguity may exist). Since the shear stress varies linearly in between the two planes, the streamwise momentum equation governing the vertical variations of the mean velocity $\overline{u}_l(z)$ integrates to $\rho_l(\nu_l+\nu_{t})d_z\overline{u}_l=G_l(z_0-z)+\rho_lu_l^{*2}$, with $G_l=-[(1-\delta_l)z_0]^{-1}\rho_lu_l^{*2}$. We introduce the relevant friction Reynolds number $Re_l^*=(1-\delta_l)z_0u_l^*/\nu_l$ and the dimensionless distances $Z=(z-\delta_lz_0)/[(1-\delta_l)z_0]$ and $\mathcal{Z}^+=(z_0-z)u_l^*/\nu_l$. With these notations, the previous momentum balance takes the dimensionless form
\begin{equation}
(1+\nu_t^+)d_Z{U}^+=Re_l^*Z\,,
\label{linstress}
\end{equation}
with $\nu_t^+=\nu_t/\nu_l$ and ${U}^+=\overline{u}_l/u_l^*$. To obtain a realistic and simple distribution of the eddy viscosity valid throughout the subsurface flow, we start from the distribution proposed in \cite{Cess1958,Reynolds1967} in the case of a fully-developed channel flow. This $\nu_t$-profile, which was designed to predict the eddy viscosity down to the wall, was later proved to be accurate even in the core of the flow \citep{Pirozzoli2014}. Disregarding the viscous contributions specific to a no-slip wall, this eddy viscosity profile reduces in present notation to
\begin{equation}
\nu_t^+=\frac{\kappa}{6}Re_l^*(1-Z^2)(1+2Z^2)\,.
\label{CRT}
\end{equation}
This expression yields $\nu_t^+\approx\kappa Re_l^*(1-Z)\equiv\kappa\mathcal{Z}^+$ in the limit $Z\rightarrow1$, with $\kappa$ the Von K\' arm\' an constant. In the core, it passes through a maximum at $Z=\frac{1}{2}$, where $\nu_t^+=\frac{3}{16}\kappa Re_l^*$, before reaching a slightly lower minimum, $\nu_t^+=\frac{1}{6}\kappa Re_l^*$, at $Z=0$.\\
Assuming $\nu_t^+\gg1$ and inserting \eqref{CRT} into \eqref{linstress} allows the latter to be integrated, yielding the velocity profile
\begin{equation}
U^+(Z)=U_0^++\frac{1}{\kappa}\log\frac{1+2Z^2}{1-Z^2}\,,
\label{velprof1}
\end{equation}
with $U_0^+=U^+(Z=0)$.
Equation \eqref{velprof1} is expected to predict reasonably well the mean velocity distribution up to the outer edge of the viscous sublayer, located approximately at $\mathcal{Z}^+=3$, i.e. $Z=1-3Re_l^{*-1}$ \citep{McLeish1975,Tsai2005,Enstad2006}. Assuming $Re_l^*\gg1$, \eqref{velprof1} predicts $U^+(\mathcal{Z}^+=3)\approx U_0^++\kappa^{-1}\log\frac{Re_l^*}{2}$. Within the viscous sublayer, \eqref{linstress} reduces to $d_Z{U}^+\approx Re_l^*$, so that
\begin{equation}
U^+(Z)=U_0^++\kappa^{-1}\log\frac{Re_l^*}{2}+3+Re_l^*(Z-1)\,.
\label{velprof2}
\end{equation}
The above model predicts a surface velocity $U^+(1)=U_0^++\kappa^{-1}\log\frac{Re_l^*}{2}+3$. With $Re_l^*=150$ and $U_0^+=0$, this yields $U^+(1)\approx13.8$, which differs by less than $8\%$ from the DNS prediction $U^+(1)\approx14.9$ of \cite{Enstad2006}. Similarly, one gets $U^+(1)-U_0^+\approx15.2$ with $Re_l^*=260$, whereas the DNS results of \cite{Tsai2005} indicate $U^+(1)-U_0^+\approx17.0$. Integrating \eqref{velprof1} from $Z=0$ to $Z=1$ yields a depth-averaged velocity $U_0^++U_{mt}$, where $U_{mt}=\kappa^{-1}(\log3-2\log2+\sqrt{2}\tan^{-1}\sqrt2)\approx2.66$. Integrating \eqref{velprof1} only up to $\mathcal{Z}^+=3$ and considering the actual velocity  profile \eqref{velprof2} within the viscous sublayer results in the viscous correction $[\frac{9}{2}-\kappa^{-1}(6\log3-3)]Re_l^{*-1}\approx-4.5Re_l^{*-1}$, leading to the slightly more accurate estimate $U_{mt}\approx2.66-4.5Re_l^{*-1}$. This finding predicts $U_{mt}\approx2.63$ for $Re_l^*=150$, which differs by $6\%$ from the value $2.48$ reported in \cite{Enstad2006}. \\
\indent The free parameters $\delta_l$ and $U_0^+$ of the above model may be used to impose approximately a zero flow rate when this constraint is appropriate. For this purpose, we consider that, in the lower part of the liquid layer extending from $z=0$ (at which a no-slip condition holds) to $z=\delta_lz_0$ (at which the shear stress vanishes), the weak remaining flow is similar to that in a half-channel with a centerline velocity $U_0^+$. In such a flow, correlations for the friction coefficient in the turbulent regime indicate that the approximate height-averaged velocity, say $U_{mb}$, is close to $0.85U_0^+$, whereas the (dimensionless) friction velocity at the corresponding wall, say $\mathcal{U}^*$, is roughly $0.05|U_0^+|$. Therefore, assuming that these correlations provide reasonable estimates despite the modest Reynolds number expected in this bottom boundary layer, the zero-flowrate condition reads approximately
\begin{equation}
(1-\delta_l)(U_0^++U_{mt})+\delta_lU_{mb}\approx(1-\delta_l)U_{mt}+(1-0.15\delta_l)U_0^+=0\,.
\label{zeroflow}
\end{equation}
Moreover, the total shear stress is linear throughout the entire liquid layer, implying
\begin{equation}
\mathcal{U}^{*2}=\frac{\delta_l}{1-\delta_l}\,,\quad\mbox{hence}\quad U_0^+\approx-20\left(\frac{\delta_l}{1-\delta_l}\right)^{1/2}\,.
\label{linshear}
\end{equation}
Solving \eqref{zeroflow}-\eqref{linshear} with the above estimate for $U_{mt}$ yields
\begin{equation}
\delta_l\approx0.017-0.06Re_{l0}^{*-1}\,,\quad U_0^+\approx2.62-4.6Re_{l0}^{*-1}\,,
\label{delU0}
\end{equation}
with $Re_{l0}=z_0u_l^*/\nu_l$ the Reynolds number based on the friction velocity $u_l^*$ at the interface and the thickness $z_0$ of the entire liquid layer.\\
\indent The above findings may straightforwardly be used to initialize the flow field in the liquid. If no condition has to be imposed on the flowrate, the bottom of the liquid layer is assumed to experience no shear and to be at rest. Therefore, one simply sets $U_0^+=0$ and $\delta_l=0$, so that the driving force per unit volume is $G_l=-\rho_lu_l^{*2}/z_0$, with the friction velocity $u_l^*$ at the liquid surface being related to that in the air flow through $u_l^*=\left(\frac{\rho_a}{\rho_l}\right)^{1/2}u^*$. This friction velocity is used to make \eqref{CRT} and \eqref{velprof1}-\eqref{velprof2} dimensional and initialize the eddy viscosity and velocity profiles with the corresponding estimates. At the bottom wall, the resolved velocity obeys a Dirichlet condition $ {\bf{\langle{u}\rangle}}={\bf{0}}$, while the eddy viscosity obeys the Neumann condition ${\bf{n}}\cdot\nabla\nu_t=0$, with ${\bf{n}}$ the normal to the wall. If the zero-flowrate condition has to be enforced, one first makes use of the estimate \eqref{delU0} for $\delta_l$ to determine $Re_l^*=(1-\delta_l)Re_{l0}^*$ and $G_l=-[(1-\delta_l)z_0]^{-1}\rho_lu_l^{*2}$. It is worth noting that the corresponding value for the driving force per unit volume is $G_l\approx-1.017\rho_lu_l^{*2}/z_0$, to be compared with $G_l=-\frac{3}{2}\rho_lu_l^{*2}/z_0$ in the laminar regime. Then, the plane $z=\delta_lz_0$ is considered as a virtual bottom sliding with speed $U_0^+u_l^*$, with $U_0^+$ as given in \eqref{delU0}. The normal velocity, $ {\bf{\langle{u}\rangle}}\cdot{\bf{n}}$, and the normal derivative of the eddy viscosity, ${\bf{n}}\cdot\nabla\nu_t$, are both assumed to vanish on that virtual bottom. Last, the $\nu_t(z)$- and $\overline{u}(z)$-profiles are initialized as above throughout the layer $\delta_lz_0<z\leq z_0$.

\section{Performance of the turbulence model in a fully-developed channel flow}
\label{channel}
\begin{figure*}[t!]
    \center
    \hspace{-7mm} \includegraphics[width=0.33\textwidth]{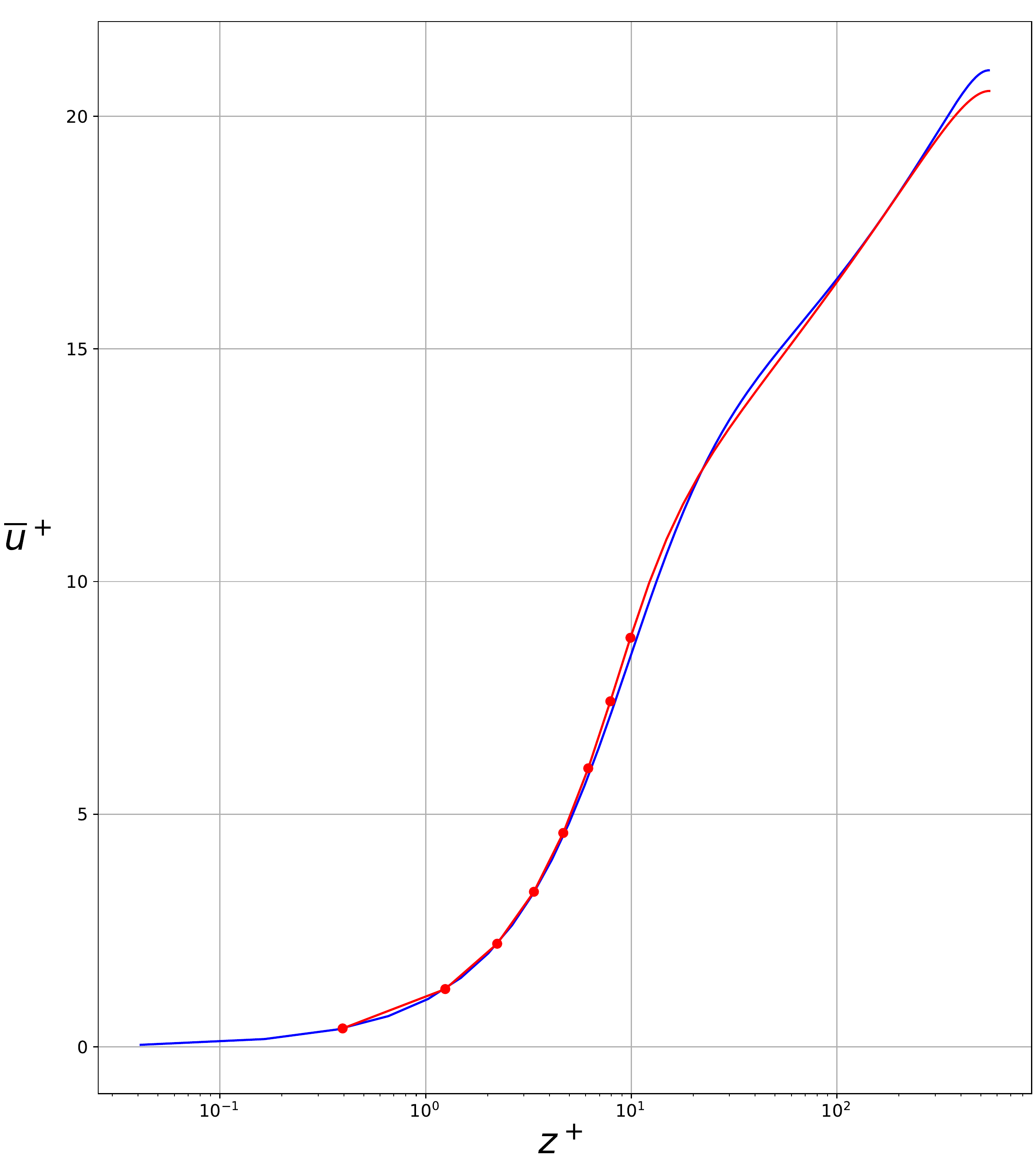}
   %    \vspace{-1mm}\hspace{10mm}$(a)$
     \hspace{-0.5mm}\includegraphics[width=0.33\textwidth]{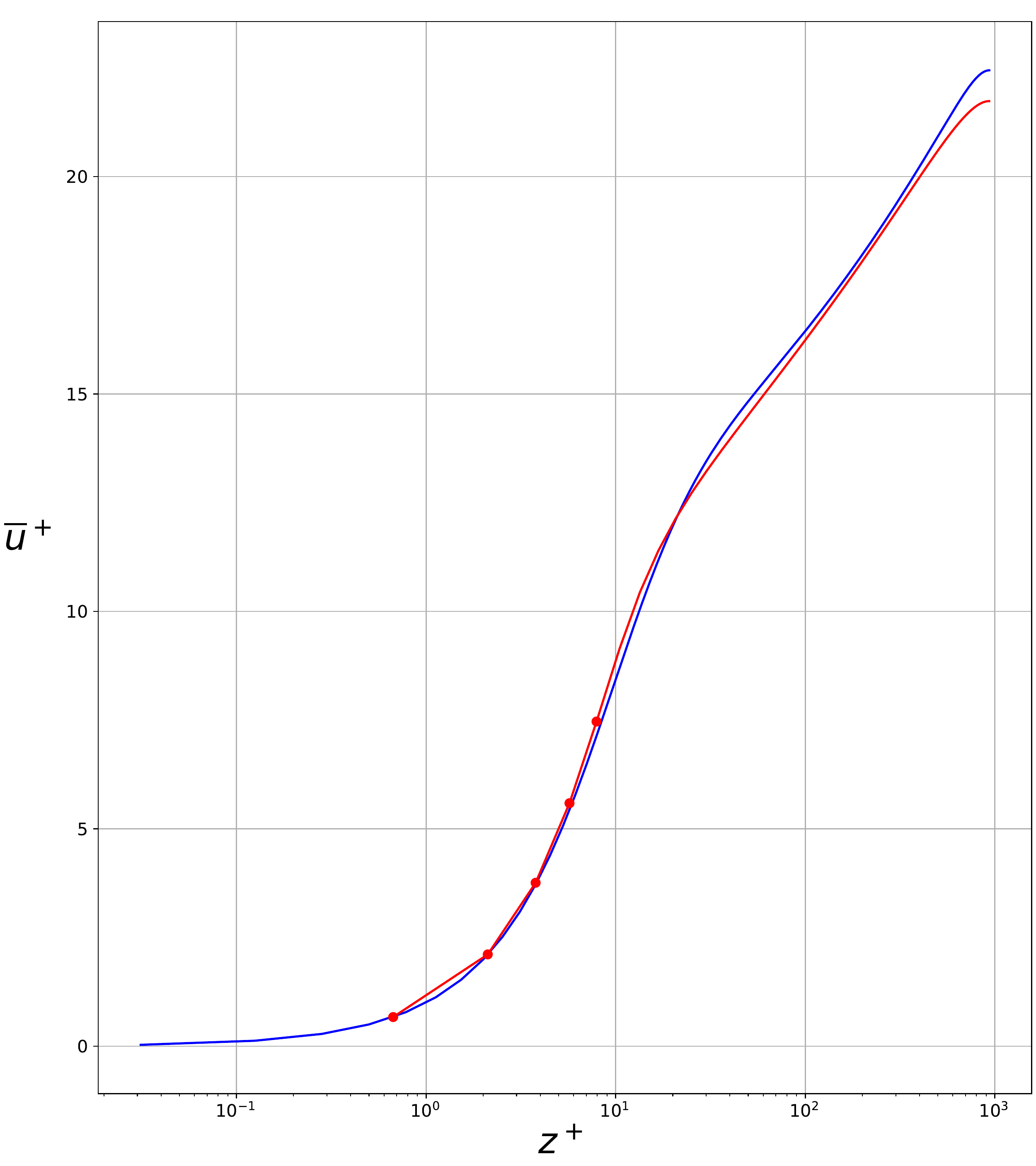}
  %    \vspace{-1mm}\hspace{10mm}$(b)$
     \hspace{-0.5mm}\ \includegraphics[width=0.33\textwidth]{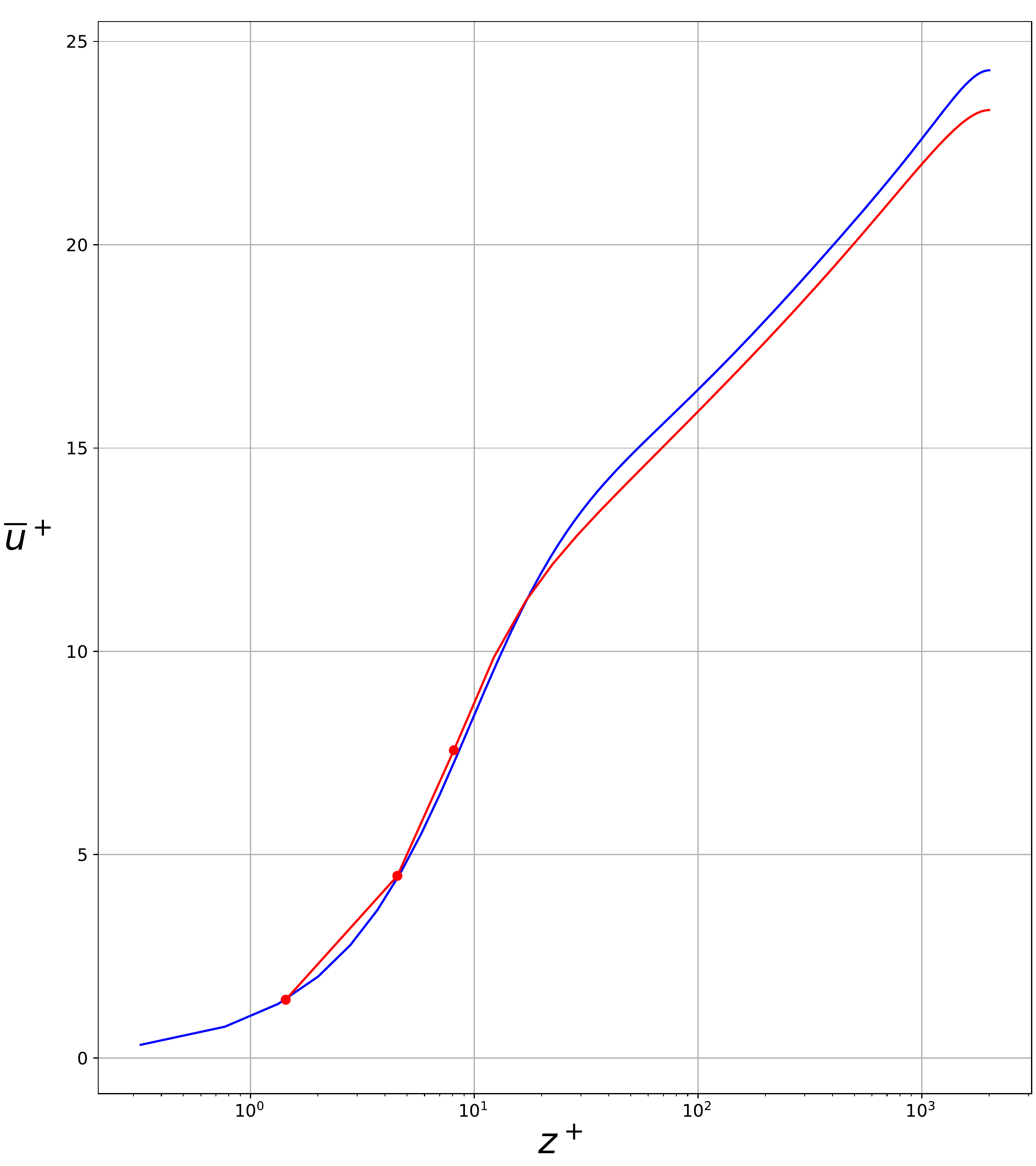}
 \vspace{-1mm}\hspace{5mm}$(a)$\hspace{50mm}$(b)$\hspace{50mm}$(c)$
    \caption{Mean velocity profiles in a plane channel at three different Reynolds numbers, all computed with the same grid. $(a)$: $Re^*=550$; $(b)$: $Re^*=934$; $(c)$: $Re^*=2003$. Red line: present results; blue line: DNS data from \cite{DelAlamo2003} ($Re^*=550$ and $934$) and \cite{Hoyas2006} ($Re^*=2003$). The red bullets indicate the grid points in the near-wall region $z^+\leq10$.}
      \label{channels}
\end{figure*}
We considered the fully-developed flow in a plane channel characterized by the friction Reynolds number $Re^*=u^*H/\nu$, with $h$ the channel half-height and $u^*$ the friction velocity related to the prescribed pressure gradient $d\mathcal{P}/dx$ through $u^*=(-\frac{1}{\rho}\frac{d\mathcal{P}}{dx}H)^{1/2}$. We simulated this flow for several Reynolds numbers up to $Re^*=2000$, to compare the mean velocity profiles with the DNS data of \cite{DelAlamo2003} and \cite{Hoyas2006} (available online at https://torroja.dmt.upm.es/channels/). For this purpose, we used a non-uniform grid with $64$ cells from the wall to the centerline. This grid was obtained by prescribing a geometric sequence with a minimum cell size $\Delta y_{min}=4.4\times10^{-3}H$ and a common ratio $1.035$. With these characteristics, the height of the cell closest to the wall ranges from $0.79\delta_\nu$ for $Re^*=550$ (figure \ref{channels})$(a)$) to $2.88\delta_\nu$ for $Re^*=2003$ (figure \ref{channels})$(c)$), with $\delta_\nu=\nu/u^*$ the characteristic near-wall length scale. We kept the grid unchanged on purpose for all $Re^*$ in order to check the sensitivity of the model to grid under-resolution in the viscous sublayer. Computations were initialized with a Poiseuille velocity half-profile corresponding to the prescribed pressure gradient and a uniform nonzero turbulent viscosity $\tnu({\bf{x}},t=0)=10\,\nu$. Figure \ref{channels} shows that the computed profiles of $\overline{u}(z)$ are in good agreement with the DNS data whatever $Re^*$. A slight underestimate of the velocity throughout the log-layer is observed at the highest $Re^*$, for which only one grid point stands within the viscous sublayer. Despite this poor near-wall resolution, the velocity profile is still closely approximated within the buffer layer, say up to $z^+\approx20$ (with $z^+=z/\delta_\nu$), and the difference between the two predictions at a given $z^+$ does not exceed $3\%$ throughout the log-layer.
\section{Influence of the relative wavelength on the performance of the turbulence model in the flow over a wavy wall}
\label{Phase}
\begin{figure}[t!]
    \center
    \hspace{-7mm} \includegraphics[width=0.5\textwidth]{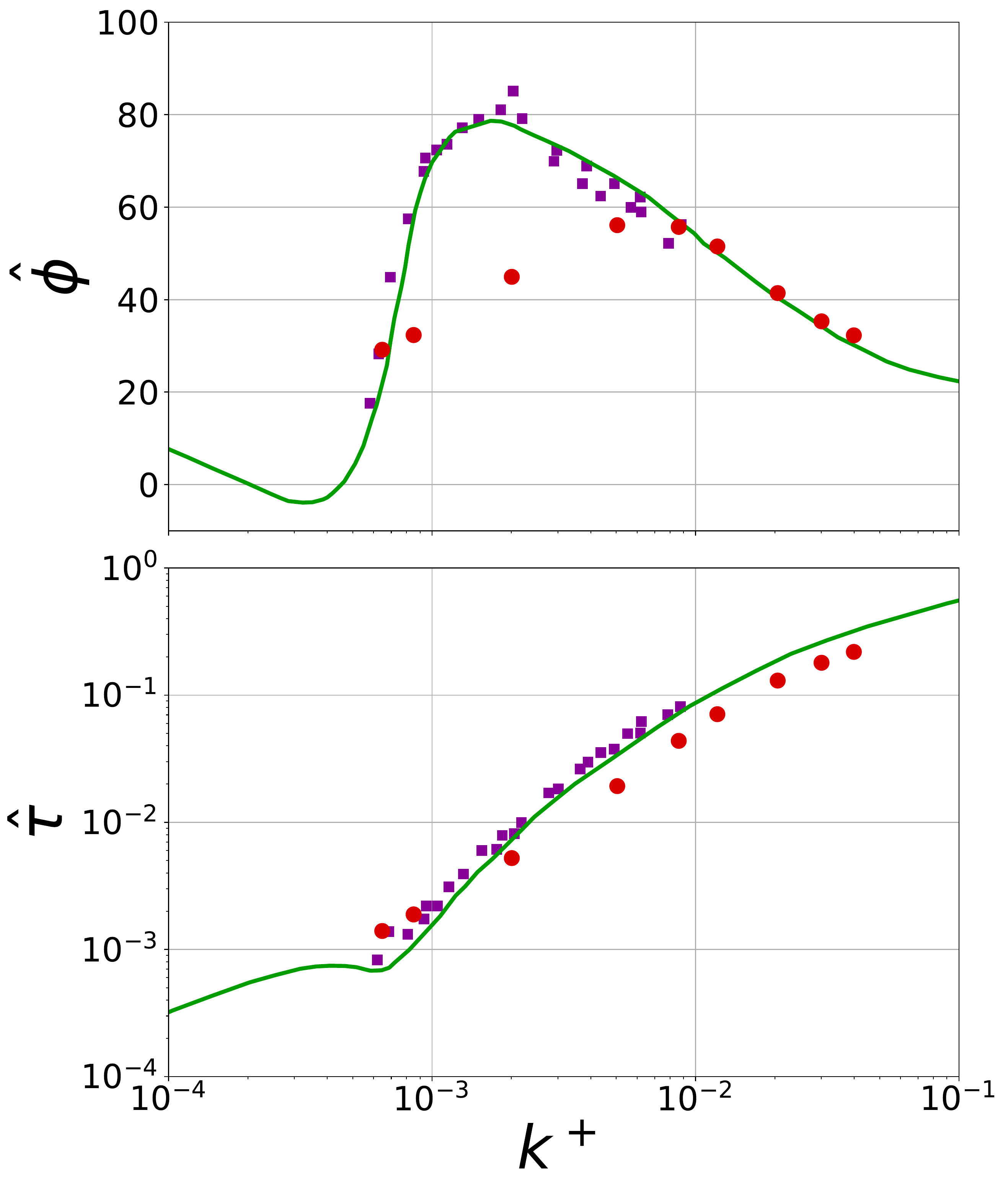}
    \caption{Variations of the phase angle (top) and amplitude (bottom) of the wall stress stress in an unseparated turbulent flow over a wavy wall.{\color{red}{\Large$\bullet$}}: present predictions for $ak=0.097$; {\color{violet}{$\blacksquare$}}: experiments from \cite{Abrams1985} with $ak=0.035$; {\color{green}{-----}}: relaxation model \citep{Abrams1985}.}
      \label{phase}
\end{figure}
In the framework of the test cases devoted to the flow over a wavy wall, we carried out several runs in which the relative wavenumber of the undulations, $k^+=k\nu/u^*$, was varied in order to quantify the range of $k^+$ over which the predictions of the turbulence model may be considered accurate. Defining the vertical position on the wavy wall as $z_w(x)=a\cos{kx}$, the wall shear stress (normalized by $\rho\overline{u^*}^2$) may be written in the form $\tau_w(x)=\overline{\tau}_w+\hat{\tau}\cos{(kx+\hat{\phi})}$. Figure \ref{phase} displays the variations of the phase angle $\hat{\phi}$ and amplitude $\hat{\tau}$ with respect to $k^+$, and compares them with the experimental data gathered in \cite{Abrams1985} and the predictions of the `relaxation' mixing length model proposed in the same reference. In this model, the Van Driest parameter routinely used to obtain accurate predictions of the mean velocity profile very close to the wall (say for $z^+<25$) is modified to account for the influence of the streamwise pressure gradient. The relevant pressure gradient at a given streamwise position is considered to have a nonzero phase shift with respect to the actual local pressure gradient. This phase shift, which results from the non-local influence of the periodic forcing imposed to the near-wall turbulence structure by the wall undulations, is estimated through a first-order differential equation modelling the corresponding relaxation. This modified mixing length model involves two additional $k^+$-independent parameters that were tuned to obtain the best possible agreement with experimental results over the whole range of $k^+$. Predictions of this relaxation model are shown with a solid line in figure \ref{phase}. It is seen that the phase angle $\hat{\phi}$ provided by the Spalart-Allmaras model agrees very well with these predictions down to $k^+\approx8\times10^{-3}$. In contrast, the Spalart-Allmaras model predicts a decrease of $\hat\phi$ with $k^+$ for $k^+\lesssim6\times10^{-3}$, while experimental data and predictions of the relaxation model show that the phase angle goes on increasing as $k^+$ decreases, down to $k^+\approx1.5\times10^{-3}$. This deficiency is actually shared by all turbulence models based on the eddy-viscosity closure and can only be removed by turning to second-order Reynolds stress models. Indeed, the eddy-viscosity concept assumes that the off-diagonal components of the Reynolds stress tensor are directly proportional to the local strain rate of the mean flow. This local relationship holds as long as the distortion introduced by the wall undulations is fast enough, but is no longer valid beyond a certain critical wavelength at which non-local effects become significant \citep{Belcher1993,Belcher1998}. It is this non-locality that was artificially introduced through the influence of the `relaxed' pressure gradient in the modified mixing length model of \cite{Abrams1985}. Fortunately, the largest wavelengths considered in the present work correspond to $k^+=5\times10^{-3}$, a position at which the one-equation model underestimates $\hat\phi$ by only $10^\circ$. Therefore, this deficiency is not expected to alter significantly the predictions of the growth of short wind-generated waves, such as those considered in \S\,\ref{evol2}. In contrast, the bottom subfigure makes it clear that the model consistently under-predicts the amplitude of shear stress variations, in line with the observations of \S\,\ref{ptest}. This is presumably its most serious limitation in the present context.   % If you have bibdatabase file and want bibtex to generate the
% bibitems, please use
%

 \bibliographystyle{elsarticle-num} 
\bibliography{turbu}

% else use the following coding to input the bibitems directly in the
% TeX file.


% \bibitem[Author(year)]{label}


% \end{thebibliography}
\end{document}